\documentclass{aastex63}
\shorttitle{Gravity, Winds, Moment of inertia}
\shortauthors{Militzer and Hubbard}
\usepackage{amsmath}
\usepackage{bm}
\newcommand{\thickhline}{%
    \noalign {\ifnum 0=`}\fi \hrule height 1pt
    \futurelet \reserved@a 
}

\newcommand{\qRot}{$q_{\rm rot}$ }
\newcommand{\QRot}{$q_{\rm rot}$}

\newcommand{\muint}{\int\limits_{-1}^{+1} d\mu}

\begin{document}

%title and author info
\title{Relation of Gravity, Winds, and the Moment of Inertia of Jupiter and Saturn}

\correspondingauthor{Burkhard Militzer}
\email{militzer@berkeley.edu}

\author[0000-0002-7092-5629]{Burkhard Militzer}
\affiliation{Department of Earth and Planetary Science, University of California, Berkeley, CA, 94720, USA}
\affiliation{Department of Astronomy, University of California, Berkeley, CA, 94720, USA}

\author{William B. Hubbard}
\affiliation{Lunar and Planetary Laboratory, University of Arizona, Tucson, AZ 85721, USA}

%Abstract - maximum 250 words for PSJ !!! - Instead of "{\em Juno}", write only "Juno"
\begin{abstract}
We study the relationship of zonal gravity coefficients, $J_{2n}$, zonal winds, and axial moment of inertia (MoI) by constructing models for the interiors of giant planets. We employ the nonperturbative concentric Maclaurin spheroid (CMS) method to construct both physical (realistic equation of state and barotropes) and abstract (small number of constant-density spheroids) interior models. We find that accurate gravity measurements of Jupiter's and Saturn's $J_2$, $J_4$, and $J_6$ by Juno and Cassini spacecrafts do not uniquely determine the MoI of either planet but do constrain it to better than 1\%. Zonal winds (or differential rotation, DR) then emerge as the leading source of uncertainty. For Saturn, they are predicted to decrease the MoI by 0.4\% because they reach a depth of $\sim$9000 km while on Jupiter, they appear to reach only $\sim$3000 km. We thus predict DR to affect Jupiter's MoI by only 0.01\%, too small by one order of magnitude to be detectable by the Juno spacecraft. We find winds primarily affect the MoI indirectly via the gravity harmonic $J_6$ while direct contributions are much smaller because the effects of pro- and retrograde winds cancel. DR contributes +6\% and $-0.8$\% to Saturn's and Jupiter's $J_6$ value, respectively. This changes the $J_6$ contribution that comes from the uniformly rotating bulk of the planet that correlates most strongly with the predicted MoI. With our physical models, we predict Jupiter's MoI to be 0.26393$\pm$0.00001. For Saturn, we predict 0.2181$\pm$0.0002, assuming a rotation period of 10:33:34 h that matches the observed polar radius. 
\end{abstract}

\keywords{giant planets, Jupiter, Saturn, interior models, gravity science}

\section{Introduction}

The angular momentum of a giant planet must be accurately known to calculate the planet's precession rate, which is the crucial quantity to determine whether it is in a spin-orbit resonance. Such resonances have been invoked, along with additional assumptions to explain the obliquities of Saturn, 27$^\circ$~\citep{Saillenfest_2021,Wisdom2022}, Jupiter, 3$^\circ$~\citep{Ward_Canup_2006}, and  Uranus, 98$^\circ$ \citep{Saillenfest_Uranus_2022}. The planetary spin angular momentum contributes 99\% of the angular momenta of the Jovian or Saturnian systems, the rest coming from the most massive satellites.  To high order, the total angular momentum of a planetary system is conserved over billions of years while the planet's moment of inertia $C$ changes \citep{Helled_2012,Nettelmann_2012} due to secular cooling and other processes like helium rain~\citep{WilsonMilitzer2012b} and core erosion~\citep{WilsonMilitzer2012}, and satellite orbits exchange angular momentum with the planet through tidal interactions \citep{Fuller_2016}. 

The space missions {\em Juno} \citep{Bolton2017} and {\em Cassini} \citep{Spilker2019} have provided us with a wealth of new data for Jupiter and Saturn. Multiple close flybys have mapped the gravity fields of these planets with a high level of precision \citep{Durante2020,Iess2019} that far exceeds the earlier measurements by the {\em Pioneer} and {\em Voyager} missions \citep{CS85,CA89}. The new measurements have also led to a revision of the assumptions that are employed when interior models are constructed. Traditionally, the interiors of Saturn and Jupiter were represented by three layer models \citep{Guillot2004,SG04,Nettelmann2012,HM16} that start with an outer layer that is predominantly composed of molecular hydrogen, a deeper layer where hydrogen is metallic, and compact core that is composed of up to 100\% of elements heavier than helium. There was sufficient flexibility in choosing the layer thicknesses and the mass fractions of helium, $Y$, and heavier elements, $Z$, to match the earlier spacecraft measurements. 

Still, the predictions from various types of three layer models were not always found to be in perfect agreement for two main reasons. Early interior models relied on the theory of figures (ToF) \citep{ZT1978}, a perturbative approach, to capture the gravitational and rotational effects in a planet's interior. Most calculations employed the third and fourth order version of the ToF but this technique has recently been extended to seventh order \citep{Nettelmann2021}. With the development of the concentric Maclaurin spheroid method (CMS), it became possible to construct giant planet interior models nonperturbatively \citep{Hubbard2013}. 

The second source of uncertainty is the equation of state (EOS) of hydrogen-helium mixtures at high pressure~\citep{Vo07,Morales2010}. While shock wave measurements \citep{Ze66} now routinely reach the relevant regime of megabar pressures \citep{Si97,Co98,Kn01,Celliers2010}, the temperatures in these experiments are much higher than those in giant planet interiors \citep{MilitzerJGR2016}. For this reason, interior models invoke theoretical methods \citep{SC95} and {\it ab initio} simulations \citep{MHVTB,NHKFRB} to construct an EOS for hydrogen-helium mixtures and then add heavy elements within the linear mixing approximation \citep{SoubiranMilitzer2015,Ni2018}. A direct experimental confirmation of the prediction from {\it ab initio} simulations of hydrogen-helium mixtures under giant planet interior conditions would be highly valuable even though simulation results for other materials were found to be in good agreement with shock experiments \citep{french-prb-09,Millot2020}. 

For Jupiter, the {\em Juno} spacecraft obtained smaller magnitudes for the harmonics $J_4$ and $J_6$ than interior models had predicted \citep{HM16}. Matching  and interpreting these measurements has led authors to introduce a number of novel assumptions into interior models. One can adopt a 1-bar temperature that is higher \citep{Wahl2017a,Miguel2022} than the {\em Galileo} value of 166.1~K or invoke a less-than-protosolar abundance of heavy Z elements \citep{HM16,Nettelmann2017,Wahl2017a}. Both modifications reduce the density of the molecular outer layer, which makes it easier to match $J_4$ and $J_6$. \citet{Wahl2017a} introduced the concept of a dilute core, which partially addressed the $J_4$-$J_6$ challenge. \citet{Debras2019} adopted the dilute core concept and then decreased the heavy Z element fraction at an intermediate layer. Most recently \citet{DiluteCore2022} matched all observed $J_n$ values exactly by simultaneously optimizing parameters of the dilute core and models for the zonal winds.

%The {\em Juno} spacecraft has collected a wealth of data in Jupiter's gravity~\citep{Iess2018}, atmospheric dynamics~\citep{Kaspi2018}, and magnetic field~\citep{Connerney2018,Moore2018}. 

%Thanks to the Jupiter and Saturn orbiters {\it Juno} and {\it Cassini}, high-precision values of the planets' zonal gravitational harmonics, $J_n$, are now available to high degree $n$, and serve as important constraints on the interior mass distributions~\citep{Durante2020,Iess2019}.  
The high-precision values from the {\it Juno} and {\it Cassini} missions for Jupiter's and Saturn's zonal gravitational harmonics, $J_n$, provide important constraints on the interior mass distributions and thereby also constrain the moment of inertia as we will demonstrate in this article. A different constraint, the value of the spin angular momentum, $\mathcal{J}$, comes from measurement of forced precession of the planet's rotation axis.  As the precession periods are very long, respectively $\sim$0.5$\times$$10^6$ years for Jupiter and $\sim$2$\times$$10^6$ years for Saturn~\citep{Ward_Canup_2006}, high-precision pole-position measurements over a long time baseline are necessary to measure $\mathcal{J}$ to better than 1\%.  In principle, if the planet rotates uniformly and its spin rate, $\omega$, is known, one can obtain the axial moment of inertia, $C$, via $C=\mathcal{J} / \omega$, which would provide an independent constraint on the interior mass distribution. 

For convenience, a planet's momentum of inertia is typically reported in normalized form, MoI$\equiv C/MR_e^2$. While we normalize by the planet's mass, $M$, and the present-day equatorial radius, $R_e$ at a pressure of 1 bar, one should note that other authors have used the volumetric radius \citep{Ni2018} or made the radius age-dependent \citep{Helled_2012}. 
%Adopting a fiducial value for the planet's equatorial radius, $R_e$, at a pressure of 1 bar, it is useful to define a dimensionless axial moment of inertia, MoI$=C/MR_e^2$, to be compared with the value for the constant-density Maclaurin spheroid MoI=2/5 (independent of rotation rate $\omega$).  
In this paper, we systematically investigate how much MoI can vary for models which have exact fixed values of $\omega$ and zonal gravitational harmonics $J_n$ up to some limiting degree $n$, and thus illustrate the role of MoI as an independent constraint.  Note that the approximate Radau-Darwin formula (e.g., \citet{RadauDarwin}), which posits a unique relation between $J_2$, $\omega$, and MoI, is too inaccurate to be relevant to this investigation because Jupiter and Saturn rotate rapidly and the density varies significantly throughout their interiors~\citep{Wahl2021}. When we construct models for giant planet interiors, we assume hydrostatic equilibrium and that the density increases with pressure. Since this concentrates mass in the planet's center, we expect the inferred MoI to be substantially less than 2/5, the value for a single constant-density Maclaurin spheroid independent of its rotation rate.

The article is organized as follows. In Sec.~\ref{methods}, we show how a planet's moment of inertia and angular momentum are calculated with the CMS method. We introduce physical and abstract models for giant planet interiors. We also explain that differential rotation (DR) in a planet has direct and indirect effects. The direct effect is introduced when the observed zonal winds, that move at different angular velocities, are projected into the interior and thereby cause a planet's angular momentum to deviate from the value of an object that rotates uniformly. However, the zonal winds also make {\em dynamical} contributions to a planet's gravitational harmonics. They thereby reduce the {\em static} contributions slightly that come from the mass distribution in the planet's interior when models are constructed to match a spacecraft's gravity measurements. This change in the mass distribution also affects the resulting moment of inertia, which we call the {\em indirect} effect of DR.  

In Sec. \ref{results}, we first discuss our predictions for Saturn's momentum of inertia and illustrate how sensitively it depends on the gravity harmonics $J_4$ and $J_6$. We find that the dynamical contributions to $J_6$ play a critical role.
Then we derive the angular momentum for arbitrary giant planets, for which the mass, equatorial radius, $J_2$, and rotational period have been measured. We present results from different models for Jupiter's interior, which includes CMS calculations that we performed for Jupiter models of other authors. Finally we compare our momentum of inertia values with earlier predictions in the literature before we conclude in Sec. \ref{conclusions}.

%A real planet's $C$, along with its 1-bar equatorial radius, tends to decrease with time as the planet cools and denser components move inward. \citet{Helled_2012} investigated the behavior of Jupiter's MoI as a function of time.  The quantity that we call MoI is equivalent to Helled's NMoI {\it at the present time}, normalizing both with the present-day $R_e$ at one-bar pressure, but MoI and NMoI diverge at earlier epochs because we normalize our MoI at all epochs with the {\it present-day} $R_e$.

\section{Methods}
\label{methods}
The normalized moment of inertia of an axially symmetric body can be derived from this integral over all fluid parcels as function of radius and $\mu=\cos(\theta)$ with $\theta$ being the colatitude,
\begin{equation}
    {\rm MoI} \equiv
    \frac{C}{MR_e^2} = \frac{2\pi}{MR_e^2} \int\limits_{-1}^{+1}d\mu \int\limits_0^{R(\mu)} dr \, r^2 \, l^2 \, \rho(r,\mu)\;, 
\end{equation}
where $l=r\sqrt{1-\mu^2}$ is the distance from the axis of rotation and $R(\mu)$ marks the outer boundary of the planet. In the CMS method, one represents the mass in the planet's interior by a series of nested constant-density spheroids each adding a small density contribution, $\delta_j$, that lets the combined density increase with depth. After carrying out the radius integration, the MoI can be written as a sum over spheroids,
\begin{equation}
    \frac{C}{MR_e^2} = \frac{2\pi}{5 MR_e^2} \sum_j \delta_j
    \int\limits_{-1}^{+1}d\mu \, r_j^5(\mu)\,
    \left[ 1-\mu^2\right]\;,
    \label{eq:MoI}
\end{equation}
where $r_j(\mu)$ marks the outer boundary of the spheroid with index $j$. For a uniformly rotating (UR) body, the normalized spin angular momentum is given by $\mathcal{J}_{\rm norm}^{\rm UR} = \sqrt{q_{\rm rot}} C / MR_e^2 $ with \qRot being the dimensionless rotational parameter,
\begin{equation}
  q_{\rm rot} = \frac{\omega^2 R_e^3}{GM}, \label{eq:qrot}
\end{equation}
that compares the magnitudes of the centrifugal and gravitational potentials.
If the body is rotating differentially, one needs to revert to the 2D integral,
\begin{equation}
    \mathcal{J}_{\rm norm}^{\rm DR} = \frac{2\pi \sqrt{q_{\rm rot}}}{MR_e^2} \int\limits_{-1}^{+1}d\mu \int\limits_0^{R(\mu)} dr \, r^2 \, l^2 \, \rho(r,\mu) \frac{v(r,\mu)}{\bar{v}(l)}\;, 
    \label{eq:J}
\end{equation}
where $v(r,\mu)$ is the fluid velocity and $\bar{v}$ is that of the uniformly rotating background, $\bar{v}= l * \omega$. For convenience, one may choose to define an effective or average moment of inertia for a differentially rotating body, 
\begin{equation}
    \bar{C}^{\rm DR}/MR_e^2 = \mathcal{J}_{\rm norm}^{\rm DR} / \sqrt{q_{\rm rot}}
    \quad,
    \label{eq:Cbar}
\end{equation}
and compare it with predictions of Eq.~\ref{eq:MoI}. We call this difference the {\em direct} effect of DR on the predicted MoI, to be compared with the {\em indirect} effect that emerged because DR affects the interior density structure and thus the calculated gravity harmonics, in particular $J_6$. In Tab.~\ref{tab:JS}, we quantify the indirect DR effect by comparing the MoI values, $C^{\rm (DR)}$ and $C^{\rm (UR)}$, derived from Eq.~\ref{eq:MoI} for a model that invokes DR effects and a model that does not when they both match the observed $J_n$.

We find that the magnitude of the direct DR effect is much smaller than the indirect one (Tab.~\ref{tab:JS}) because contributions from pro- and retrograde jets to the direct effect partially cancel. Direct DR effects increase Jupiter's MoI by 0.0015\% because the prograde winds in the equatorial region dominate. For Saturn, we find that the retrograde winds at a latitude of $\sim$35$^\circ$ dominate over the prograde equatorial jet, which implies that direct DR effects lower the planet's angular momentum by $-0.13\%$.

\subsection{CMS Technique}

The spheroid surfaces $r_j(\mu)$ are contours of constant pressure, temperature, composition, and potential. The potential combines centrifugal and gravitational contributions, $Q+V$. According to \citet{ZT1978}, the gravitational potential can be expanded in the following form,
\begin{equation}
  V(r,\mu) = \frac{GM}{r} \left[ 1 - \sum_{n=1}^{\infty} \left(R_e/r\right)^{2n} J_{2n}P_{2n}(\mu) \right]
               \quad,
\end{equation}
where $P_n$ are the Legendre polynomials of order $n$ and the $J_n$ are the gravity harmonics given by
\begin{equation}
  J_n = - \frac{2 \pi}{M R_e^n} \int\limits_{-1}^{+1} d \mu \int\limits_0^{R(\mu)} \!\!\! dr \,\, r^{n+2} \,\, P_n(\mu) \,\, \rho(r,\mu)
  \label{standard_J}\quad.
\end{equation}
According to \citet{Hubbard2013}, the gravitational potential $V_j$ of a point $(r_j,\mu)$ on spheroid $j$ is decomposed into contributions from interior spheroids ($j=i \ldots N-1)$,
\begin{equation}
    V_i^{\rm (int)} (r_i,\mu) = -\frac{GM}{r_i} \sum^{N-1}_{j=i} \sum^\infty_{n=0} 
      J_{j,n} \left( \frac{R_e}{r_i} \right)^n P_{n}(\mu)
\end{equation}
and exterior spheroids ($j=0 \ldots i-1)$,
\begin{equation}
    V_i^{\rm (ext)} (r_i,\mu) = -\frac{GM}{r_i} 
         \sum^{i-1}_{j=0} \left[ J''_{j,0} \left(\frac{r_i}{R_e}\right)^3 +
    \sum^\infty_{n=0} J'_{j,n} \left( \frac{r_i}{R_e} \right)^{n+1}
    P_n(\mu)  
    \right].
\end{equation}
Following the derivation in \citet{Hubbard2013}, we define the interior harmonics
\begin{equation}
    J_{i,n} = -\frac{1}{n+3} \, \frac{2 \pi }{M} \, \delta_i \, \muint\ \, P_n(\mu) \, \left(\frac{r_i}{R_e}\right)^{n+3}
    \label{eq:harmonics}
\end{equation}
and the exterior harmonics
\begin{equation}
   J'_{i,n} = - \frac{1}{2-n} \, \frac{2 \pi}{M} \, \delta_i
   \muint \, P_n(\mu) \, \left(\frac{r_i}{R_e}\right)^{2-n}
   \label{eq:harmonics_prime}
\end{equation}
with a special case for $n=2$,
\begin{equation}
  J'_{i,n} = - \frac{2 \pi}{M} \, \delta_i 
    \muint \, P_n(\mu) \, \log\left(\frac{r_i}{R_e}\right) 
    \label{eq:harmonics_prime_n=2}
  \end{equation}
and finally,
\begin{equation}
  J''_{i,0} = \frac{2\pi\delta_i a^3}{3 M},
  \label{eq:J_prime_prime}
\end{equation}
where $M$ is the total mass of the planet. One should note that during the numerical evaluation of these expressions, it is recommended to work with harmonics that have been renormalized by the powers of the equatorial spheroid radii, $\lambda_i$. These equatorial points $(r_j=\lambda_j,\mu=0)$ serve as anchors for all spheroid shapes. This is where the reference value of the potential is computed that one uses to adjust the spheroid shape until a self-consistent solution emerges for which all spheroids are equipotential surfaces. 

It is important to choose the $\lambda_i$ grid points wisely in order to minimize the discretization error that is inherent to the CMS approach. We recommended choosing them so that a logarithmic grid in density emerges, $\rho(\lambda_{i+1})/\rho(\lambda_i)$=constant~\citep{Militzer2019a}. %, which is also the preferred choice for other numerical methods~\citep{galerkin}.
This grid choice allows us to obtain converged results when we construct our physical models with $N_S=2048$ spheroids.

In addition to gravity, one needs to consider the centrifugal potential, which takes the following simple form for a uniformly rotating body, $Q(l) = \frac{1}{2} l^2 \omega^2$. We employ this formula when we construct models for Jupiter's interior and then introduce DR effects by solving the thermal wind equation~\citep{Kaspi2016} to derive the density perturbation, $\rho'$,
\begin{equation}
\frac{\partial \rho'}{\partial s} = \frac{2 \omega}{g}\frac{\partial}{\partial z}\left[\rho u\right]
\quad,
\end{equation}
for a rotating, oblate planet~\citep{Cao2017} in geostrophic balance. $z$ is the vertical coordinate that is parallel to the axis of rotation. $\rho$ is static background density that we derive with the CMS method. $u$ is the differential flow velocity with respect to the uniform rotation rate, $\omega$. $g$ is the acceleration that we derive from the gravitational-centrifugal potential, $V+Q$, in our CMS calculations. $s$ is the distance from the equatorial plane along a path on an equipotential. We represent the flow field $u$ as a product of the surface winds, $u_s$, from \citet{Tollefson2017} and a decay function of $\sin^2(x)$ form from \citet{Militzer2019a}. This function facilitates a rather sharp drop similar to functions employed in \citet{Galanti2021} and \citet{Dietrich2021}.

Since the winds on Saturn reach much deeper, we treat them nonperturbatively by introducing DR on cylinders directly into the CMS calculations by modifying the centrifugal potential,
\begin{equation}
Q(l) = \int_0^l dl' \; l' \; \omega(l')^2
\end{equation} 
Since we assume potential theory, a cylinder's angular velocity, $\omega(l)$, cannot decay with depth, which means we are only able to include the prograde equatorial jet and first retrograde jet at $\sim$35$^\circ$ that were characterized by tracking the cloud motion in Saturn's visible atmosphere~\citep{SL2000,GM2011}.

\subsection{Physical Interior Models}

\begin{figure}
\plotone{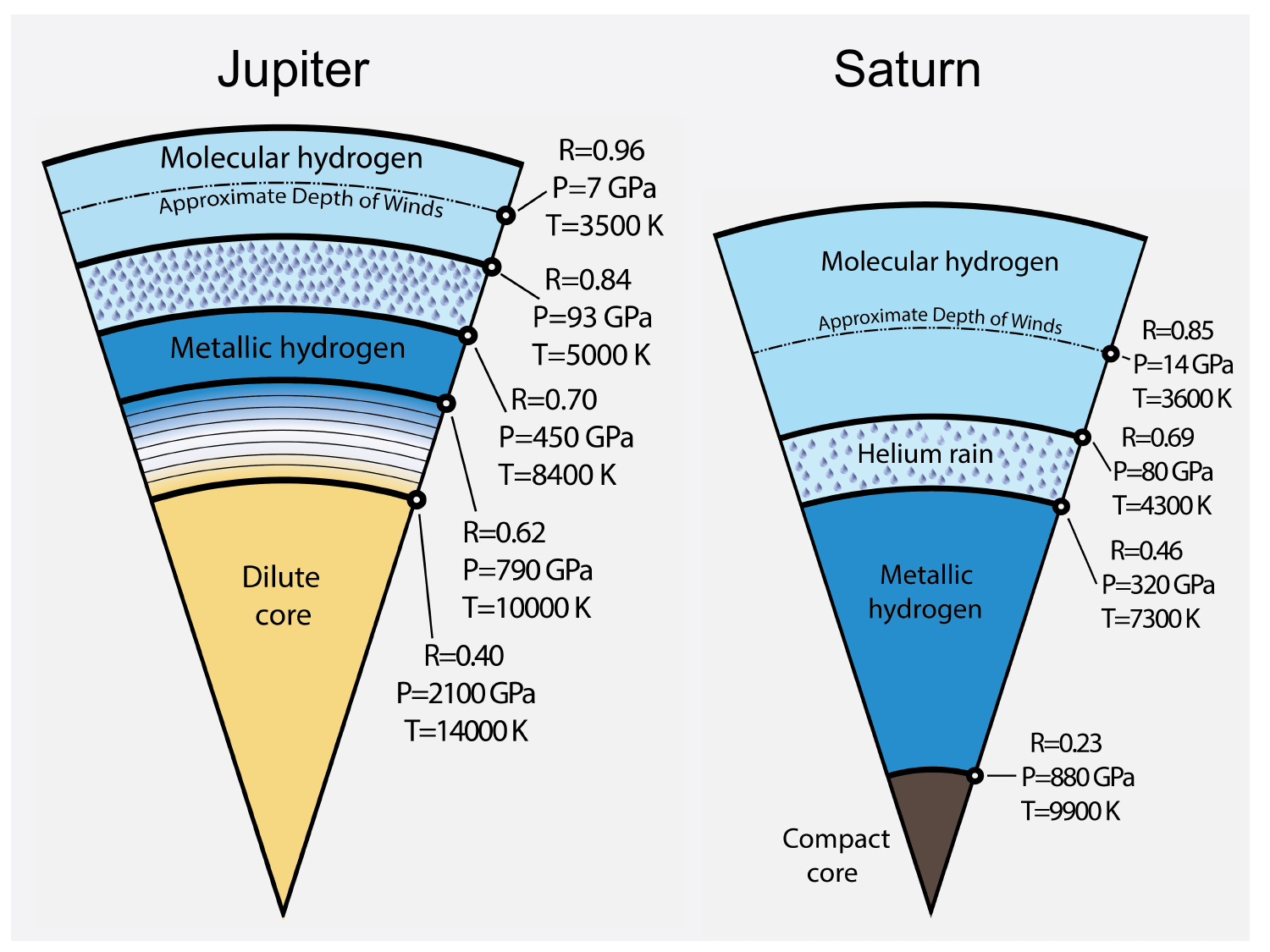}
%\gridline{\fig{Jupiter_odd_and_even_Jn_32.pdf}{0.7\textwidth}{}}
\caption{Models of Jupiter and Saturn based on CMS calculations that match the gravity measurements of both planets. The zonal winds on Saturn are predicted to reach a depth of $\sim$9000 km, involving $\sim$7\% of the planet's mass. On Jupiter, they are predicted to reach $\sim$3000 km and thus involve only 1\% of the planet's mass. Because Jupiter is more massive, the pressure rises more rapidly with depth. Therefore the helium rain layer, predicted to start approximately at 80--100~GPa, is located closer to the surface. While the gravity measurements for Jupiter imply that the planet has a dilute core, the state of Saturn's core is less certain. Here we show a model with a compact core constructed to match the gravity measurements.
\label{fig:JandS}}
\end{figure}

In Fig.~\ref{fig:JandS}, we illustrate our physical interior models for Jupiter and Saturn. Since planets cool by convection, we assume most layers in their interiors are isentropic and of constant composition. We represent their outer envelope where hydrogen is molecular by the parameters ($S_{\rm mol}$, $\tilde{Y}_{\rm mol}$, $Z_{\rm mol}$) for entropy, helium mass fraction and the fraction of heavy elements. We define $\tilde{Y} \equiv Y/(X+Y)$ with $X$ and $Y$ being the mass fractions of hydrogen and helium so that $X+Y+Z=1$. We require $Z_{\rm mol}$ to be at least protosolar, $Z_{\rm PS}=1.53\%$~\citep{Lodders2010}. The entropy is chosen to match the temperature at 1 bar: 142.7~K for Saturn~\citep{Lindal1981} and 166.1~K for Jupiter~\citep{Seiff1998} that was measured {\em in situ} by the {\em Galileo} entry probe. For Jupiter, we also consider an alternate, slightly higher temperature of 170~K from a recent reassessment of the {\em Voyager} radio occultation measurements~\citep{Gupta_2022}. 

To construct EOSs for models in this article, we start from the {\em ab initio} EOS that \citet{MH13} computed for one hydrogen-helium mixing ratio. With these calculations, absolute entropies \citep{Militzer2013} were derived that implicitly set the temperature profiles in our models. We use our helium EOS from \citet{Mi06,Mi09} to perturb helium fraction in our H-He EOS as we detailed in \citet{HubbardMilitzer2016}. We also follow this article when we introduce heavily elements into our models. Their detailed composition is not important as long as they are substantially more dense than hydrogen and helium. Ice, rocky materials and iron are all sufficiently dense so that they add mass but do not increase the volume of the mixture too much. At low pressure where the {\em ab initio} simulations do not work, we revert back to the semi-analytical EOS by \citet{Saumon1995}.

When hydrogen assumes an atomic/metallic state at approximately 80--100 GPa~\citep{Morales2009}, helium remains an insulator and the two fluids are predicted to become immiscible~\citep{stevenson-astropj-77-i,Brygoo2021}. There is indeed good evidence that helium rain has occurred in Jupiter because the {\it Galileo} entry probe measured a helium mass fraction of $\tilde{Y}=0.238\pm0.005$~\citep{vonzahn-jgr-98} that is well below the protosolar value of 0.2777~\citep{Lodders2010}. Furthermore, neon in Jupiter’s atmosphere was measured to be nine-fold depleted relative to solar, and this can be attributed to efficient dissolution in helium droplets~\citep{Roulston1995,Wilson2010}. So for our Jupiter models, we adopt the value from the {\it Galileo} entry probe for $\tilde{Y}_{\rm mol}$ and for Saturn, we make it a free parameter but constrain it to be no higher than the protosolar value because we have no information on how much helium rain has occurred in this planet. 

For both planets, we chose values for the beginning and ending pressures of the helium rain layer that are compatible with the immiscibility region that \citet{Morales2013} derived with {\it ab initio} computer simulations (see \citet{Militzer2019a} for details). Across this layer, we assume $(S,\tilde{Y},Z)$ vary gradually with increasing pressure until they reach the values of the metallic layer ($S_{\rm met}$, $\tilde{Y}_{\rm met}$, $Z_{\rm met}$) where they are again constant since we assume this layer to be homogeneous and convective. $\tilde{Y}_{\rm met}$ is adjusted iteratively so that the planet as a whole assumes a protosolar helium abundance. This also assures $\tilde{Y}_{\rm met} > \tilde{Y}_{\rm mol}$. We prevent the heavy element abundance from decreasing with depth, $Z_{\rm met} \ge Z_{\rm mol}$. Every layer is either homogeneous and convective or Ledoux stable~\citep{Ledoux1947}. This sets our models apart from those constructed by \citet{Debras2021} who introduced a layer where $Z$ decreases with depth in order to match Jupiter's $J_4$. Instead our Jupiter models all have a dilute core with $Z \approx 0.18$ (see Fig.~\ref{fig:JandS}) because this key restriction allows us to match the entire set of gravity measurements of the {\em Juno} spacecraft under one set of physical assumptions~\citep{DiluteCore2022}.

For our Monte Carlo calculations of Jupiter's interior \citep{Militzer_QMC_2023}, we vary the beginning and end pressure of the helium rain layer but apply constraints so that they remain compatible with H-He phase diagram as derived by \citet{Morales2009}. We also vary a parameter $\alpha$ that controls the shape of the helium profile in this layer, as we explain in \citet{DiluteCore2022}. During the Monte Carlo calculations, we also vary the beginning and end pressure of the core transition layer, which we assume to be stably stratified since the abundance of heavy elements increases from $Z_{\rm mol}$ to $Z_{\rm met}$ in this layer. We also allow $Z_{\rm mol}$ and $Z_{\rm met}$ to vary as long as they meet the constraint we discussed in the previous paragraph. More details of our Monte Carlo approach are given in \citet{DiluteCore2022}.

For our Saturn models, we assume a traditional compact core that is composed up to 100\% of heavy elements because this assumption was sufficient to match the gravity measurements by the spacecraft~\citep{Iess2019}, but there are alternate core models constructed to match ring seismological data~\citep{Mankovich2021}. 

\subsection{Abstract $N$ Spheroid Models}
In the previous section, we described physical interior models in hydrostatic equilibrium that rely on a realistic EOS for H-He mixtures. To explore more general behavior, we now investigate simplified models with $N_S$ spheroids. We still require each spheroid surface to be an equipotential but spheroid densities, $\rho_i$, are arbitrary as long as the densities monotonically increase toward the planet's interior, $\rho_{i+1} \ge \rho_i$. We can set the density of the outermost spheroid to zero, $\rho_0=0$, because in realistic interior models, the density of the outermost layer is typically much lower than that of deeper layers. (We also construct models in which $\rho_0$ is a free parameter, but they behave similarly, and in the limit of large $N_S$, the difference becomes negligible.)

We initialize the equatorial radii of all spheroids, starting from $i=0 \ldots N_S-1$, to fall in a linear grid, $\lambda_i = 1-i/N_S$. While we keep the outermost spheroid anchored at $\lambda_0=1$, we repeatedly scale all interior $\lambda_{i>0}$ points uniformly to obtain a model that matches the planet's mass and $J_2$ exactly. We add a penalty term to the Monte Carlo (MC) cost function if $\lambda_{i} > \lambda_0$. 

Since matching $M$ and $J_2$ requires two free parameters, we also scale all density values, $\rho_i$, uniformly. So after every update of the spheroid shapes, we employ a Newton-Raphson step to scale $\rho_i$ and $\lambda_i$ grids simultaneously. We also institute a maximum density of 10 PU (planetary unit of density, $M/R_e^3$) to prevent pathological situations in which the radius of the innermost spheroid becomes very small while its density becomes extremely large. \citet{Movshovitz_2020} and \citet{Neuenschwander_2021} also introduced upper limits on density. 
We consider 10 PU to be a reasonable choice because for Jupiter, it corresponds to a density of 52~g$\,$cm$^{-3}$, which exceeds the density of iron that is $\sim$$27$~g$\,$cm$^{-3}$ at Jupiter's core conditions~\citep{WilsonMilitzer2014}. The described set of assumptions lead to a stable procedure with $N_S$--1 free input parameters ($\rho_{i>0}$) that is amendable for MC sampling. 

Since we do not employ a physical EOS or make specific assumptions about the planet's composition or temperature profile, our abstract models share similarities with the empirical models by \citet{Helled_2009} and \citet{Neuenschwander_2021} or the composition-free models by \citet{Movshovitz_2020} who represented the Saturn interior density profile by three quadratic functions before conducting MC calculations to match the {\em Cassini} gravity measurements.

%%%%%%%%%%%%%%%%%%%%%%%%%%%%%%%%%%%%%%%%%%%%%%%%%%%%%%%%%%%%%%%%%%%%%%%%%%%%%%%%%%%%%%%%
%%%%%%%%%%%%%%%%%%%%%%%%%%%%%%%%%%%%%%%%%%%%%%%%%%%%%%%%%%%%%%%%%%%%%%%%%%%%%%%%%%%%%%%%
%%%%%%%%%%%%%%%%%%%%%%%%%%%%%%%%%%%%%%%%%%%%%%%%%%%%%%%%%%%%%%%%%%%%%%%%%%%%%%%%%%%%%%%%

\section{Results}
\label{results}
\subsection{Saturn}

\begin{figure}
\plotone{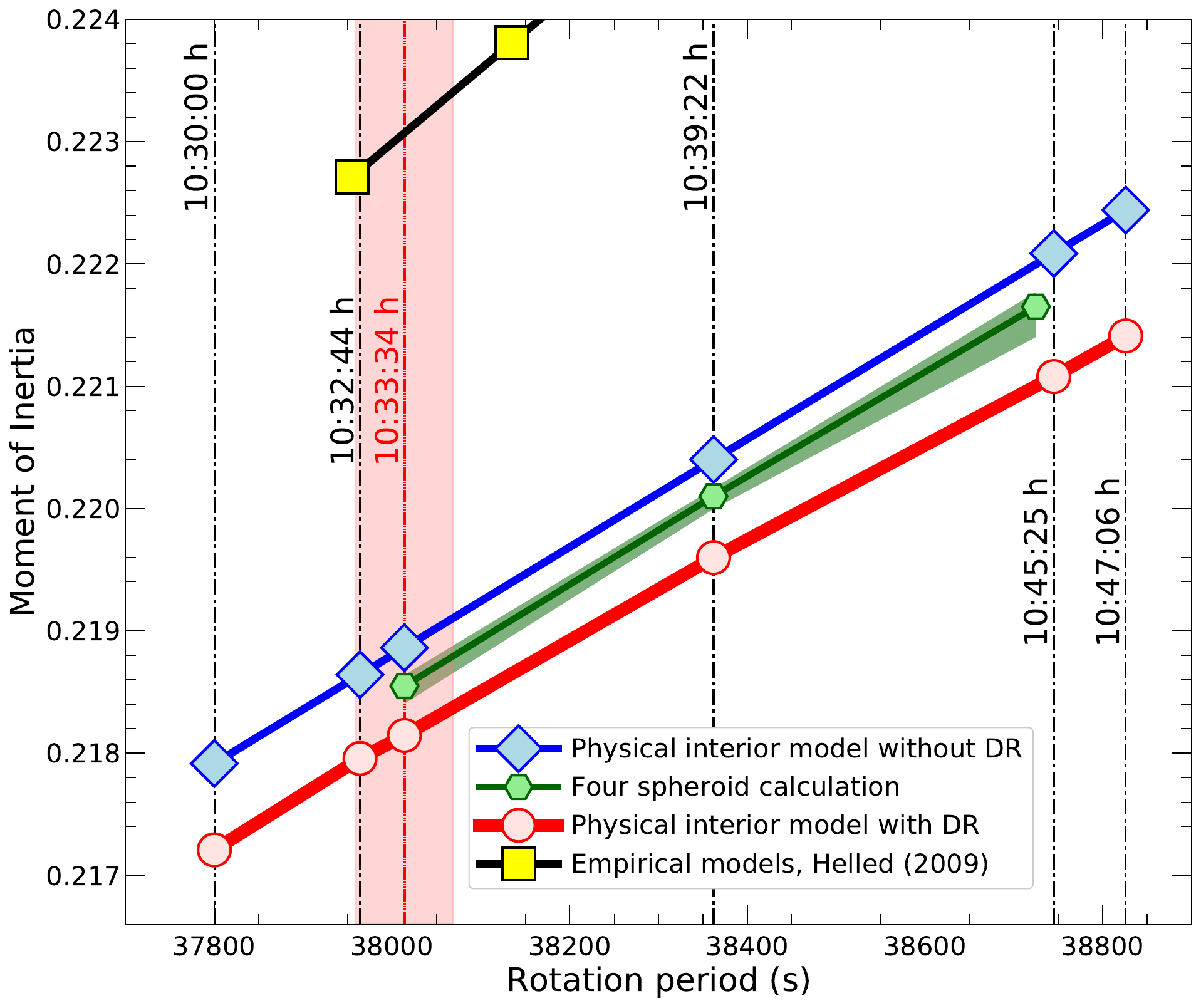}
%\gridline{\fig{Jupiter_odd_and_even_Jn_32.pdf}{0.7\textwidth}{}}
\caption{MoI of Saturn computed for different rotation periods that have been assumed in the literature.  The physical model with differential rotation (DR) matches the measured gravity harmonics $J_2$-$J_{12}$ while models without DR can only match values up to $J_6$. We prefer the period of 10:33:34~hr~$\pm$~55~s because it allows models with DR to match the {\em Voyager} measurements of the planet's polar radius. Under these assumptions, we predict Saturn's MoI $= 0.2181 \pm$~0.0002. With a low-order theory of figures, \citet{Helled_2009} predicted Saturn's MoI to be 2\% larger (yellow squares). The green band illustrates the range of predictions with four-spheroid calculations that were reported by \citet{Wisdom2022}.
\label{fig:Saturn}}
\end{figure}

\begin{figure}
%\plotone{J4_J6_plot_17.png}
\gridline{\fig{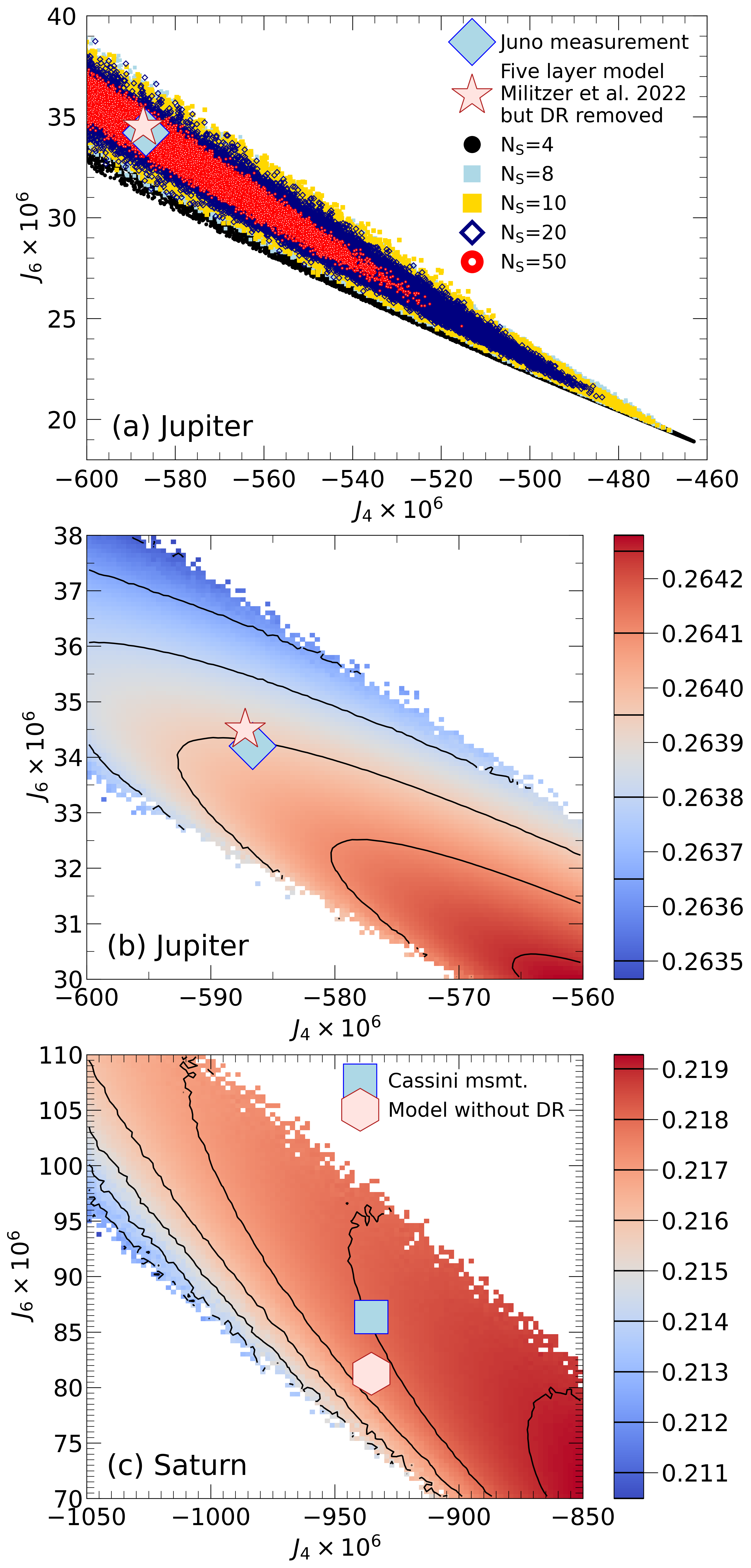}{0.45\textwidth}{}}
\caption{ Posterior distributions of Monte Carlo ensembles of abstract models for Jupiter's and Saturn's interior without DR that match the observed values for \qRot and $J_2$. The blue symbols represent measurements of {\em Juno} and {\em Cassini} spacecraft while the red symbols show predictions from models without DR effects. In panel (a), we compare ensembles of models with various numbers of spheroids. Counterintuitively, models with fewer spheroids tend to show a wider range of $J_4$ and $J_6$ values (see text). In panels (b) and (c), the background color, the color bar and the contour lines represent the average MoI as function of $J_4$ and $J_6$. $N_S=20$ spheroids were employed. DR effects alter Saturn's MoI significantly while they are less important for Jupiter. Jupiter's MoI decreases with rising $J_6$ while it is almost independent of $J_4$. Conversely, Saturn's MoI strongly depends on $J_4$ but still increases with rising $J_6$. 
\label{fig:J4J6points}}
\end{figure}

\begin{figure}
\plotone{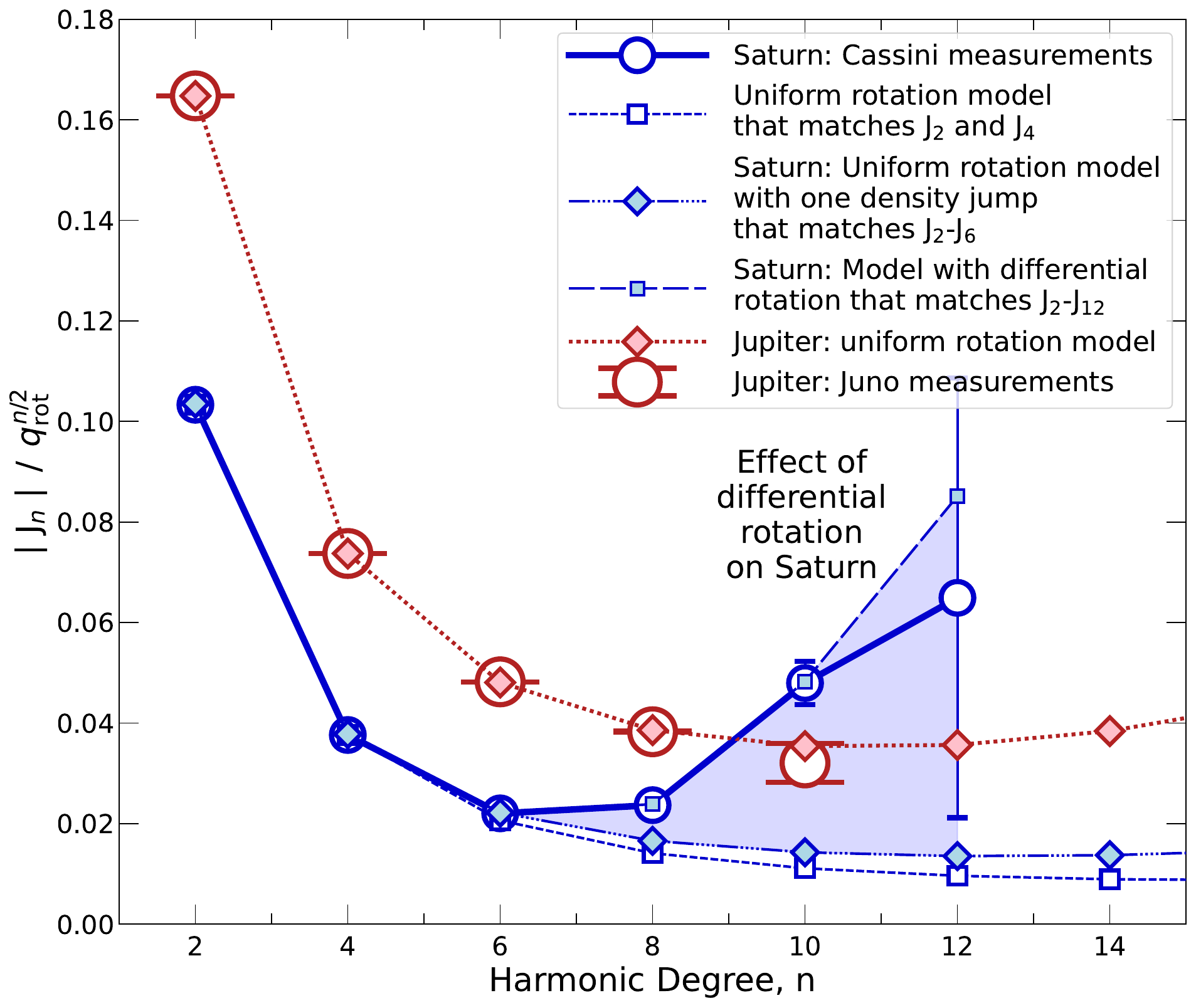}
%\gridline{\fig{Jupiter_odd_and_even_Jn_32.pdf}{0.7\textwidth}{}}
\caption{The even gravitational moments $J_n$ of Jupiter and Saturn versus degree $n$. 
% $q_{\rm rot}=\Omega^2R_e^3/GM$,
All moments have been scaled by powers of the rotational parameter, \qRot, because the ratio $J_n / q_{\rm rot}^{n/2}$ is approximately constant according to the theory of figures (ToF)~\citep{ZT1978}. The variation of the ratios with the order $n$ is due to the contribution of higher-order terms in the ToF, captured to high precision in the nonperturbative CMS method. The figure also illustrates that the effects of differential rotation are much stronger for Saturn (blue shaded area) than for Jupiter because Saturn's winds extend to a greater depth of $\sim$9000~km~\citep{Iess2018}. Saturn models with differential rotation fit the observed moments up to $J_{12}$ while uniform rotation models can only fit coefficients up to $J_4$, or up to $J_6$ if a density jump is included. Here we compare the {\em Juno} measurements with a uniform rotation model for Jupiter's interior while the model with differential rotation in~\citet{DiluteCore2022} matches the entire set of gravity coefficients.
\label{fig:Jn}}
\end{figure}

%Saturn [\qRot=0.1576653506, $J_2$=0.016290573, MoI=0.21796] 
%Jupiter [\qRot=0.08919543238, $J_2$=0.0146965063, MoI=0.263929] 
\begin{table}
\centering                                                        
\begin{tabular}{lll}
\toprule
 & Jupiter & Saturn \\
\hline
 GM [$10^{16}$ m$^3$s$^{-2}$] & 12.66865341$^\dagger$ & 3.7931208\\
 Equatorial radius, $R_e$, at 1 bar [km] & 71492 & 60268\\
 Measured $J_2 \times 10^6$ &    14696.5063 $\pm$ 0.0006$^\dagger$  &    16324.108 $\pm$ 0.028$^*$\\
 Measured $J_4 \times 10^6$ &  ~~--586.6085 $\pm$ 0.0008$^\dagger$  &  ~~--939.169 $\pm$ 0.037$^*$\\
 Measured $J_6 \times 10^6$ &  ~~~~~34.2007 $\pm$ 0.0022$^\dagger$  &  ~~~~~86.874 $\pm$ 0.087$^*$\\
 %Calculated $\Delta J_2^{\rm wind} \times 10^6$ &  ~~~~~~~0.0579$^\ddag$ & $\pm 50^\Box$\\
 %Calculated $\Delta J_4^{\rm wind} \times 10^6$ &  ~~~~~~~0.2377$^\ddag$ & $-30 \ldots 10^\Box$\\
 %Calculated $\Delta J_6^{\rm wind} \times 10^6$ & ~~~~~~--0.2684$^\ddag$ & $\sim4^\diamondsuit$\\
 Period of rotation & 9:55:29.711 h & 10:33:34 h~$\pm$ 55~s  \\
 \hline
 Inferred \QRot, Eq.~(\ref{eq:qrot}) & 0.08919543238 & 0.1576653506 \\
 Calculated ratio of volumetric and equatorial radii, $R_m/R_e$ &  0.97764461 & 0.96500505\\
% $1-(R_m/R_e)^2$ &  0.044211008 & 0.068765250\\
 Calculated MoI, $C/MR_e^2$, Eq.~(\ref{eq:MoI}) & 0.26393~$\pm$~0.00001 & 0.2181~$\pm$~0.0002\\
 Calculated angular momentum, $\mathcal{J_{\rm norm}}$, Eq.~(\ref{eq:J}) & 0.078826~$\pm$~0.000003 & 0.08655~$\pm$~0.00008\\
 Direct DR effect, $(\bar{C}^{\rm DR}-C)/C$, Eqs.~(\ref{eq:MoI},\ref{eq:Cbar})& +1.5$\times 10^{-5}$ & --1.3$\times 10^{-3}$ \\ 
 Indirect DR effect, $(C^{\rm (DR)}-C^{\rm (UR)})/C^{\rm (DR)}$, Eq. (\ref{eq:MoI}) & --1$\times 10^{-4}$ & --3$\times 10^{-3}$ \\
\hline
\end{tabular}
\caption{Parameters for Jupiter and Saturn that we used for this article. $^*$Measurements from \citet{Iess2019} but converted to our 1~bar radius. $^\dagger$value and 1-$\sigma$ uncertainty from \citet{Durante2020}. 
%$^\ddag$ from \citet{DiluteCore2022}. $^\diamondsuit$estimated from \citet{Iess2019} $^\Box$ wide range of Monte Carlo ensemble from \citet{Galanti2019}.
\label{tab:JS}}
\end{table}

In Fig.~\ref{fig:Saturn}, we show MoI values computed for the physical models of Saturn's interior in Fig.~\ref{fig:JandS}, as well as for the abstract $N_S$ spheroid models. The dominant source of uncertainty in the computed MoI is the planet's period of rotation, which cannot be derived from the planet's virtually axisymmetric magnetic field. This is not the case for Jupiter, whose rotation period is known to a fraction of a second (see Tab.~\ref{tab:JS}). Without any constraints on the rotation period, the predictions for Saturn's MoI vary by $\sim$2\%. Still all values predict that Saturn is not currently in a spin-orbit resonance with Neptune today~\citep{Wisdom2022}. For all rotation periods shown in Fig.~\ref{fig:Saturn}, we can construct interior models that match the entire set of gravity coefficients that the {\em Cassini} spacecraft measured during its ultimate set of orbits~\citep{Iess2019}, so gravity measurements alone are insufficient to constrain the rotation period. Only if we match the planet's polar radius as measured by the {\em Voyager} spacecraft using radio occultation, the now-preferred period of 10:33:34 h~$\pm$ 55~s emerges~\citep{Militzer2019a}.
This rotation period is in remarkably good
agreement with the value of 10:33:38$\,$h$^{+112\rm s}_{~-89 \rm s}$ inferred from waves observed in Saturn's rings \citep{Mankovich2019}.

Once a rotation period has been selected, the remaining uncertainty is dominated by effects of differential rotation (DR), which amount to about 0.4\%. Without DR effects, we are only able to match the gravity harmonics $J_2$--$J_6$, and already matching $J_6$ requires us to introduce one additional adjustable parameter, so we add an artificial density jump ~\citep{Iess2019}. The comparison of predictions from model with and without DR in Fig.~\ref{fig:Jn} illustrates that DR effects are much more important for Saturn than for Jupiter. When we include DR effects in our Saturn models, we are able to match the entire set of gravity harmonics $J_2$--$J_{12}$ without an artificial density jump. We find that resulting MoI drops 0.4\% below predictions from models that match $J_2$--$J_6$ without DR. % MORE WORK WOULD BE NEEDED TO KEEP THE FOLLOWING: This drop can be attributed to the role of the gravity harmonic $J_6$. As we will show later when we discuss Jupiter, $J_6$ and the MoI are negatively correlated. In~\citet{Iess2019}, it was shown that physical models without DR yield $J_6$ values between 80.7 and 81.7 $\times 10^{-6}$ while models with DR can match the measured $J_6$ value of (86.340~$\pm$~0.087) $\times 10^{-6}$ exactly. This means that approximately 5 $\times 10^{-6}$ 

To better understand this drop, we constructed MC ensembles of abstract models of Saturn's interior that match \qRot and $J_2$ without invoking DR. In Fig.~\ref{fig:J4J6points}c, we plot the posterior distribution of the computed MoI in $J_4$-$J_6$ space. We also show the {\em Cassini} measurements~\citep{Iess2019} and the model from Fig.~\ref{fig:Jn} without DR nor artificial density jump, matching the observed $J_2$ and $J_4$. We estimate DR effects increase Saturn's $J_6$ from $\sim$$81 \times 10^{-6}$ to the observed value of 86.340 $\times 10^{-6}$. Fig.~\ref{fig:J4J6points}c shows that the {\em Cassini} measurements place Saturn in a regime where an increase in $J_6$ (or in $J_4$) leads to an increase in the MoI: $\frac{\partial \rm MoI}{\partial J_6^{\rm int}}>0$. 

At the same time, models without DR in Fig.~\ref{fig:Saturn} predict a larger MoI than models with DR. This lets us conclude that when models with DR are constructed to match the {\em Cassini} measurements, DR effects reduce the contribution to $J_6$ that comes from the uniformly rotating bulk of the interior, $J_6^{\rm int}$. So when models with and without DR are compared, both matching the spacecraft data, models with DR predict a {\it smaller} MoI because their $J_6^{\rm int}$ is reduced by contributions to $J_6$ from DR. It is primarily this change to the $J_6$ term that affects the MoI while the DR contributions to $J_2$ and $J_4$ are too small to matter. On the other hand, DR effects dominate the higher order $J_n$ starting with $J_8$ (see Fig.~\ref{fig:Jn}) but their values are controlled by the outer layers of the planet~\citep{Guillot2005,N13,Fortney2016,MilitzerJGR2016} where the density is comparatively low, and therefore they do not contribute much to the MoI. We conclude that DR effects couple to the MoI mostly via $J_6$.

\begin{figure}
%\plotone{MoI_histogram_04.pdf}
\gridline{\fig{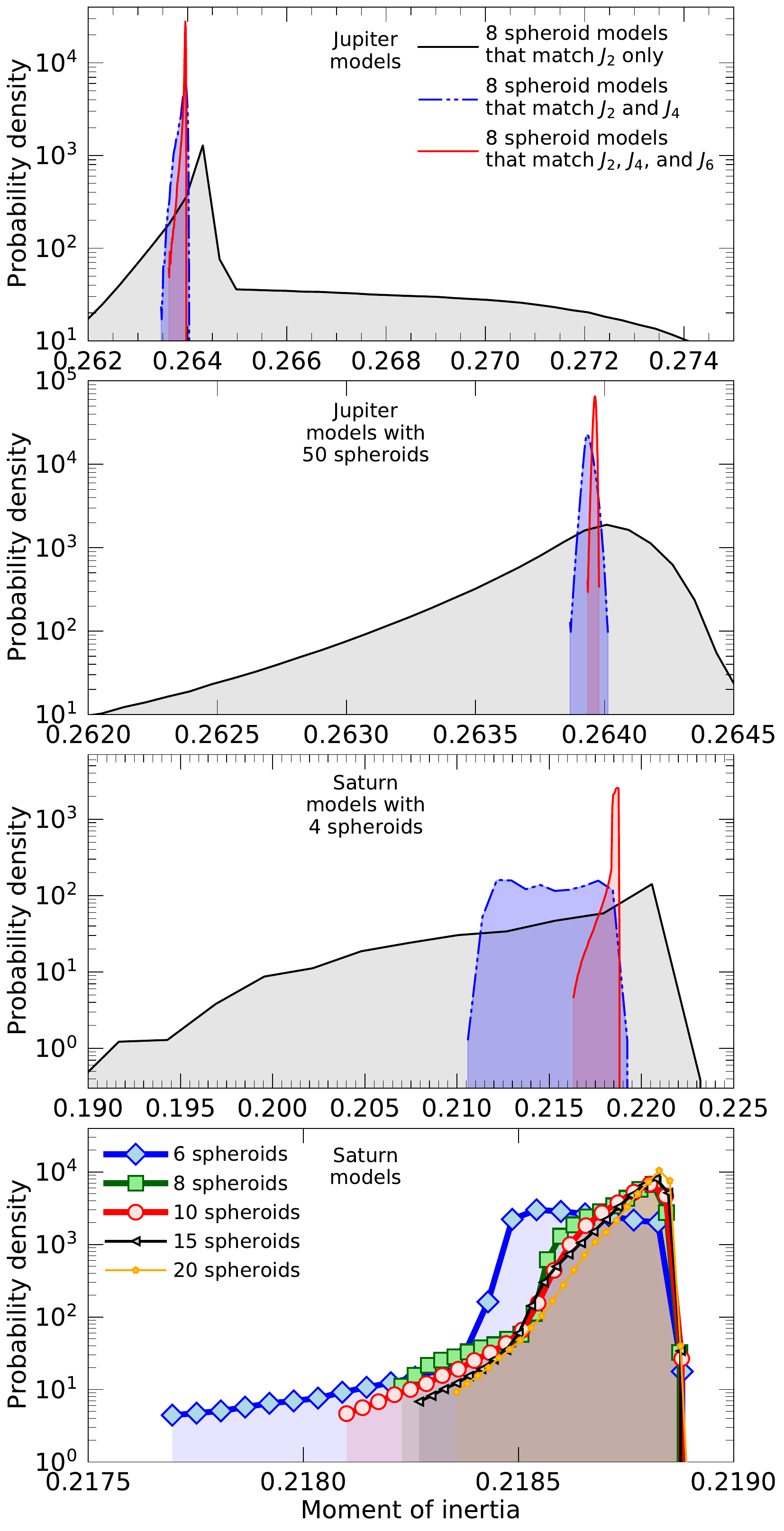}{0.5\textwidth}{}}
\caption{ Probability density distributions of Jupiter's and Saturn's  MoI values. The top panel illustrates how the range  of likely MoI values shrinks as models with eight constant-density spheroids are required to first match only Jupiter's $J_2$ value, then to match  $J_2$ and $J_4$, then to reproduce the measured values for $J_2$ through $J_6$.  The second panel shows that the range of likely MoI values shrinks further when this calculation is instead performed with 50 constant-density  spheroids (see Fig.~\ref{fig:J4J6points}). The third panel shows the  same trend in four-spheroid models of Saturn. The lowest panel shows predictions from models that match Saturn's $J_2$ through $J_6$.  As the number of spheroids is increased, models tend to cluster in a narrower MoI interval. With 50 spheroid models that match $J_2$ through $J_6$, we obtained a range from 0.26393--0.26398 for Jupiter's MoI. (Between $5 \times 10^5$ and $6 \times 10^7$ models were constructed to compute every individual MoI histogram.)
\label{fig:J2J4J6}}
\end{figure}

While the models in Fig.~\ref{fig:J4J6points} only match $J_2$, we compare MC ensembles of Saturn models in Fig.~\ref{fig:J2J4J6} that either match $J_2$ and $J_4$ or all three $J_2$--$J_6$. The posterior distribution of MoI value narrows substantially with every additional constraint. %This figure also show that the range of predicted MoI value tends to shrink if more spheroids 

Abstract models that match $J_2$--$J_6$ yield a MoI range from $\sim$0.2180 until a sharp drop off at 0.2189. Our physical models yield a MoI value of 0.2181 with a 1-$\sigma$ error bar of 0.0002. Broadly speaking the predictions from the two ensembles are compatible. However, with increasing spheroid number, our abstract models cluster around the most likely value of 0.2188, which is a bit higher than our physical models predict. This difference is a consequence of the way the two ensembles are constructed. In one case, we apply a number of physical assumptions. In the other, we do not and let the Monte Carlo procedure gravitate towards the most likely parameter space as long as the spacecraft measurements are reproduced. So one may expect to see small deviations in the predictions of the two ensembles.

%%%%%%%%%%%%%%%%%%%%%%%%%%%%%%%%%%%%%%%%%%%%%%%%%%%%%%%%%%%%%%%%%%%%%%%%%%%%%%%%%%%%%%%%
%%%%%%%%%%%%%%%%%%%%%%%%%%%%%%%%%%%%%%%%%%%%%%%%%%%%%%%%%%%%%%%%%%%%%%%%%%%%%%%%%%%%%%%%
%%%%%%%%%%%%%%%%%%%%%%%%%%%%%%%%%%%%%%%%%%%%%%%%%%%%%%%%%%%%%%%%%%%%%%%%%%%%%%%%%%%%%%%%

\subsection{Giant planets in general}

\begin{figure}
\plotone{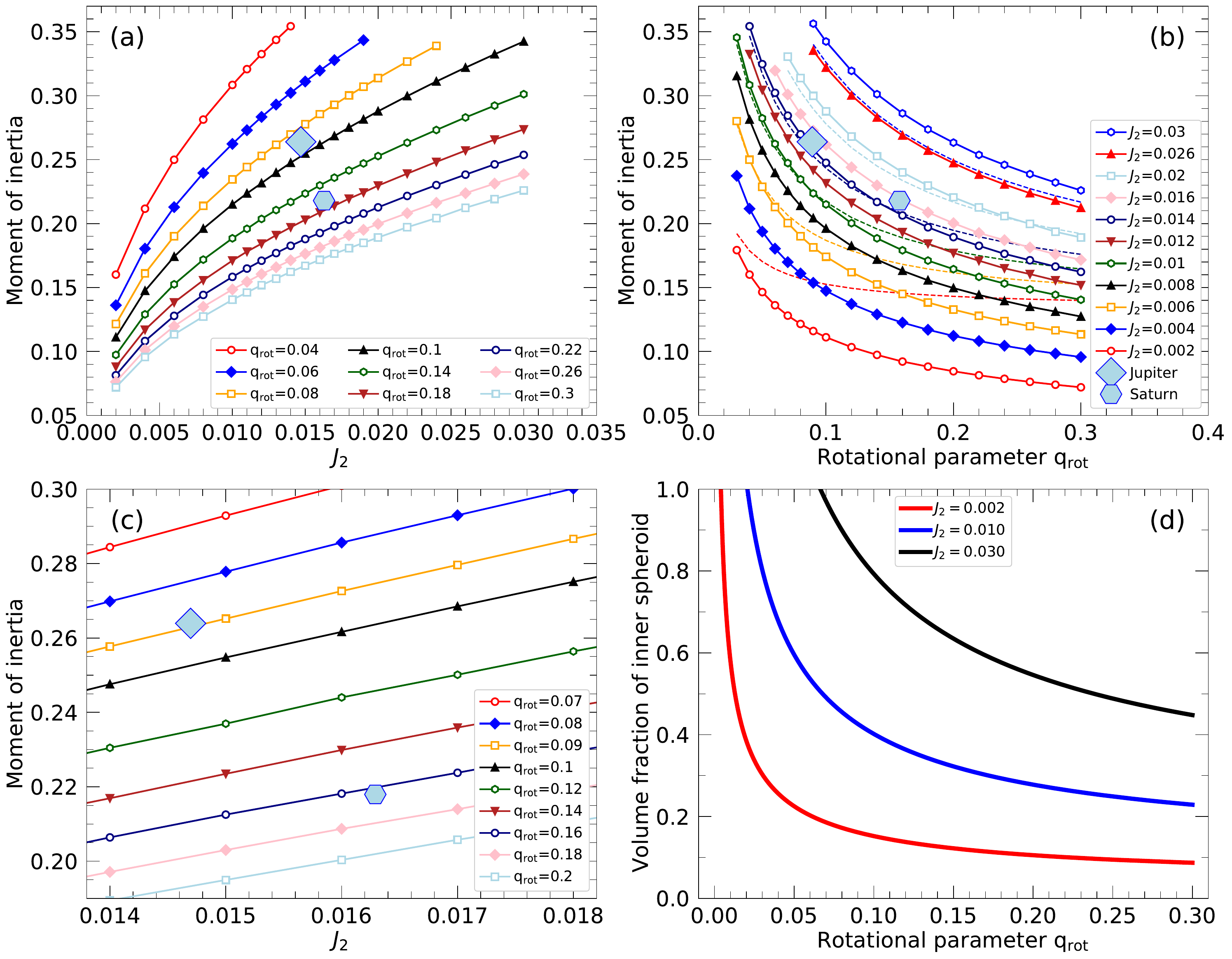}
\caption{Panels a), b), and c) show the MoI of hypothetical planets with prescribed \qRot and $J_2$ values. The solid lines show the ensemble average of abstract models with 50 constant-density spheroids. The predictions from physical giant interior models that match the spacecraft measurements of Jupiter and Saturn are shown for comparison. The dashed lines in panel b) are the prediction from the Radau-Darwin Eq.~\ref{Darwin-Radau} that becomes exact in the limit of small \qRot and large $J_2$. In this limit, the 50-spheroid models are forced to approach the uniform-density limit of MoI=$\frac{2}{5}$. One can also approach this limit with two-spheroid calculations. So in panel d), we show how the volume fraction of the inner spheroid changes as a function of \QRot. As the fraction approaches unity, the choices for \qRot and $J_2$ imply a constant planet. 
\label{MoI:qRotJ2}}
\end{figure}

The results in Fig.~\ref{fig:J2J4J6} show that the MoI of a giant planet can already be constrained reasonably well even if only \qRot and $J_2$ are known. We therefore derive the MoI for a set of hypothetical giant planets by performing MC calculations with $N_S=50$ spheroids on a grid of \qRot and $J_2$ points, which will help us to understand why Jupiter's and Saturn's MoI differ by $\sim$20\%. 

The ensemble averages of the computed MoI are shown in Fig.~\ref{MoI:qRotJ2}. One finds in Fig.~\ref{MoI:qRotJ2}a that for a given \QRot, the MoI rises rapidly with increasing $J_2$. To first approximation, $J_2$ is a measure of the planet's oblateness. So if $J_2$ is increased, while the equatorial radius and the rotation period are kept constant, more mass is moved towards the equatorial region, increasing the MoI. 
In Fig.~\ref{MoI:qRotJ2}b, we also show the predictions of the Radau-Darwin approximation,
\begin{equation}
    {\rm MoI} = \frac{2}{3} \left(1- \frac{2}{5} \sqrt{\eta-1} \right)
    \;\; {\rm with} \;\; \eta = \frac{5 q_{\rm rot}}{3 J_2 + q_{\rm rot}}
    \label{Darwin-Radau}
\end{equation}
While there exist slightly different formulations of this approximation~\citep{ZT1978}, they all become exact in the limit of small \qRot and large $J_2$. In this limit, the planet's density becomes more and more uniform throughout its interior. Eventually the MoI approaches $\frac{2}{5}$, the value for a uniform-density fluid planet (Maclaurin spheroid) regardless of rotation rate. The $\frac{2}{5}$ value cannot be exceeded unless one permits the density in the interior to be less than that of the exterior, which we exclude from consideration. 

The uniform-density limit is also approached by models that have just two spheroids. While we fix the parameters of the outer spheroid, $\rho_0=0$ and $\lambda_0=1$, the two parameters of the inner spheroids, $\rho_1 \ge 0$ and $\lambda_1 \le 1$, are just sufficient to match a pair of prescribed \qRot and $J_2$ values. In Fig.~\ref{MoI:qRotJ2}d, we plot the volume fraction of the inner spheroid as function of \QRot. When this fraction approaches 1 for small \QRot, the density of the planet becomes uniform. For a given $J_2$, this occurs at the same \qRot value that leads to a MoI value of $\frac{2}{5}$ in Fig.~\ref{MoI:qRotJ2}b. The two-spheroid calculations in Fig.~\ref{MoI:qRotJ2}d also confirm the trends that we see in the $N_S$ spheroid calculations in the other figure panels: With increasing \QRot, more and more mass needs to be concentrated in the planet's center to satisfy the $J_2$ constraint. This leads to a decrease in the MoI if \qRot is increased for a given $J_2$, explaining the trends in Fig.~\ref{MoI:qRotJ2}b.

% If one increases \qRot while keeping $J_2$ constant, one finds the MoI decreases, as mass is concentrated near the axis of rotation to yield the same $J_2$ value for planets spinning at an increasing rate. 

%Jupiter = [ 0.08919543238, 0.0146965063,     0.26392903   ] 
%Saturn  = [0.1576653506,   0.01629057300000, 0.2179557945 ]
%{\bf In Fig.~\ref{MoI:qRotJ2}, we also show the calculated MoI values of Saturn and Jupiter, which differ by $\sim$20\%. Fig.~\ref{MoI:qRotJ2} shows that this difference in magnitude is primarily the consequence of the differing \qRot value. For fixed \QRot, the deviation between the $J_2$ values of the two planets would yield only a 5\% difference in the predicted MoI. } {\it This paragraph is subject to misinterpretation and I recommend that we remove it.}

Finally we performed calculations for our two-spheroid models for Saturn's and Jupiter's \qRot and $J_2$ values. While such models are crude, they show that the volume fraction of the inner spheroid is $\sim$54\% for Jupiter and only $\sim$42\% for Saturn. This implies that a higher fraction of Saturn's mass is concentrated near the the center, consistent with the fact that typical Jupiter models have a dilute core, while Saturn models matching the gravity measurements typically do not require one. 

%%%%%%%%%%%%%%%%%%%%%%%%%%%%%%%%%%%%%%%%%%%%%%%%%%%%%%%%%%%%%%%%%%%%%%%%%%%%%%%%%%%%%%%%
%%%%%%%%%%%%%%%%%%%%%%%%%%%%%%%%%%%%%%%%%%%%%%%%%%%%%%%%%%%%%%%%%%%%%%%%%%%%%%%%%%%%%%%%
%%%%%%%%%%%%%%%%%%%%%%%%%%%%%%%%%%%%%%%%%%%%%%%%%%%%%%%%%%%%%%%%%%%%%%%%%%%%%%%%%%%%%%%%

\subsection{Jupiter}

In Fig.~\ref{fig:J4J6points}a, we compare the posterior distributions of abstract Jupiter models with different numbers of spheroids. All models were constructed to match Jupiter mass, equatorial radius, and $J_2$ exactly. Models with fewer spheroids tend to show a wider range of $J_4$ and $J_6$ values, which is counterintuitive because, e.g., the entire space of 10 spheroid models is included in that of the 20 spheroid models. (In an 20 spheroid model, one only needs to set $\rho_{2i}=\rho_{2i+1}$ to obtain a valid 10 spheroid model.) However, the available space of 20 spheroid models is much bigger and in most models, the magnitude of the density steps, $\rho_{2i} \le \rho_{2i+1}$, is smaller than that between two densities in a 10 spheroid model. In most 20 spheroid models, the density varies slightly more gradually than in the coarser 10 spheroid models. As a result, a representative set of 20 spheroid models occupies a smaller area in $J_4$-$J_6$ space than a set of 10 spheroid models.  Despite this reduction with increasing $N_S$, the range of every model ensemble includes the $J_4$ and $J_6$ values from the {\em Juno} measurements~\citep{Durante2020} as well as the predictions from the static gravity terms (no DR) according to the dilute core models from~\citet{DiluteCore2022}. We will refer to them as five layer models throughout this article. 

In Fig.~\ref{fig:J4J6points}b, we compare the average MoI as function of $J_4$ and $J_6$. In general, small $J_6$ and $J_4$, that are less negative, lead to larger MoI values. One also notices that as $J_6$ is increased for a given $J_4$, the MoI goes through a maximum and the {\em Juno} measurements place Jupiter in the regime where $\frac{\partial \rm MoI}{\partial J_6^{\rm int}}<0$ while the opposite is true for Saturn. From the shape of contour lines, we can infer that Jupiter's MoI is almost independent of $J_4$. 

The five layer models from \citet{DiluteCore2022} predict DR contributions to Jupiter's $J_6$ to be negative: --0.27 $\times 10^{-6}$ or --0.8\%. They are much smaller in magnitude than for Saturn (it was +6\%) and have the opposite sign. % see **7315**
However, since $\frac{\partial \rm MoI}{\partial J_6^{\rm int}}$ also has the opposite sign, we are again in a situation where models matching the gravity data with DR effects predict a smaller MoI than models without DR. The magnitude of the MoI difference between the two types of models is, at --0.01\%, much smaller for Jupiter while it was --0.4\% for Saturn. 

While a --0.01\% correction was derived from our more recent five layer models~\citep{DiluteCore2022}, one may also ask whether the DR effect could make a larger contribution to $J_6$. Our preliminary Jupiter model~\citep{HubbardMilitzer2016}, put together before {\em Juno} data became available, differs in $J_6$ by --0.8 $\times 10^{-6}$ from the now-available gravity data. Even if such a large discrepancy came from DR effects, the MoI would only decrease by --0.04\%, still smaller than the 0.1\% precision that {\em Juno} is expected to ultimately achieve for the MoI measurements.

\begin{figure}
\plotone{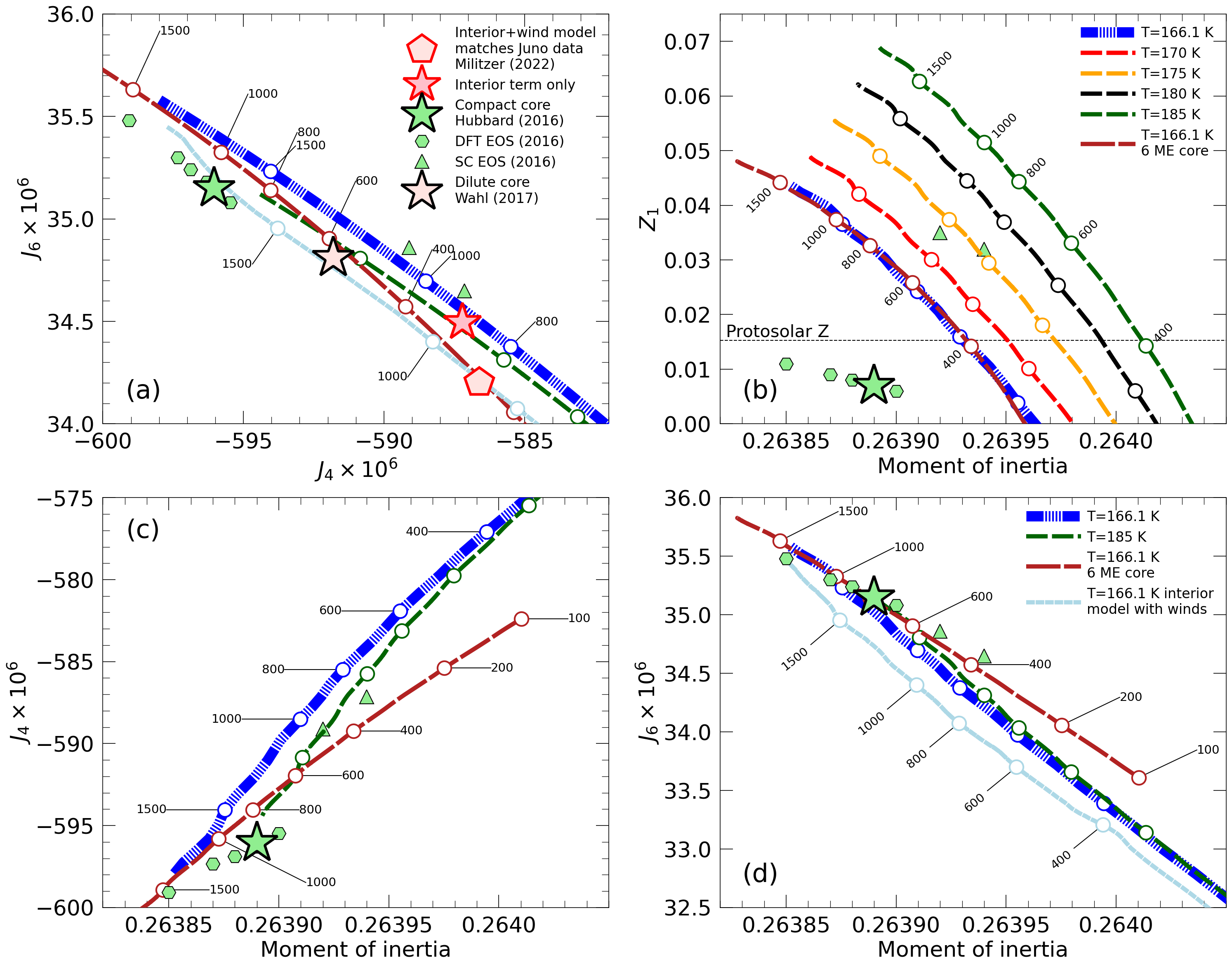}
%\gridline{\fig{MoI_histogram_06.pdf}{0.6\textwidth}{}}
\caption{ Predictions from two and three layer models for Jupiter's interior that were constructed under different assumptions and match $J_2$ exactly. Colors label models consistently across all panels but not all curves are shown in every panel for clarity. The numbers specify the assumed transition pressure in GPa between molecular and metallic layers so that models can be traced across different panels. Panel (b) compare the abundance of heavy elements in the outer molecular layer, $Z_1$, with the value of the protosolar nebula, 1.53\%~\citep{Lodders2010}. No winds were included except for the last models shown in light blue color. Only the brown curves refer to models with a compact core (6 Earth masses, rocky composition). The light green star shows the preferred model from \citet{HubbardMilitzer2016} while the pentagons and triangles indicate other compact core models based on the EOSs by \citet{MH13} and \citet{SC95}.
\label{fig:23layers}}
\end{figure}

In Fig.~\ref{fig:23layers}, we compare the MoI of two and three layer models for Jupiter's interior \citep{saumon-apj-04,Guillot2004,MHVTB} that are based on a physical EOS for the hydrogen-helium mixture but do not contain sufficient flexibility to match all observations. The predicted MoI values range from 0.26385--0.26400. In panel \ref{fig:23layers}b, the temperature of Jupiter's interior was increased by raising the 1 bar temperature step by step from the value of the Galileo entry probe, 166.1~K, up to the extreme value of 185~K~\citep{Miguel2022}. Raising 1 bar temperature lowers the density of the hydrogen-helium mixture, which enables one to add more heavy elements and thereby produce models that have at least a protosolar heavy element abundance, $Z_{\rm PS}=1.53\%$. An increase of 10~K allows one to approximately add one $Z_{\rm PS}$ worth of heavy elements to an existing model. Still most models require the transition pressure to be 400~GPa or higher, which is not compatible with predictions for the metallization of hydrogen and for the hydrogen-helium immiscibility. Both are assumed to occur at approximately 80--100 GPa~\citep{Morales2010}.

Like the abstract models in Fig.~\ref{fig:J4J6points}, all physical models in Fig.~\ref{fig:23layers} match $J_2$ exactly but the fact that the equation of hydrostatic equilibrium is satisfied and that a physical EOS is employed means that $J_4$ and $J_6$ are now much more tightly correlated. While abstract models permitted a wide interval of $J_6$ values from 32.5--36.5$\times 10^{-6}$ for $J_4=-587 \times 10^{-6}$, the more physical assumptions narrow this range to 34.2--34.5$\times 10^{-6}$ in Fig.~\ref{fig:23layers}a.

\begin{figure}
%\plotone{MoI_histogram_07.pdf}
\gridline{\fig{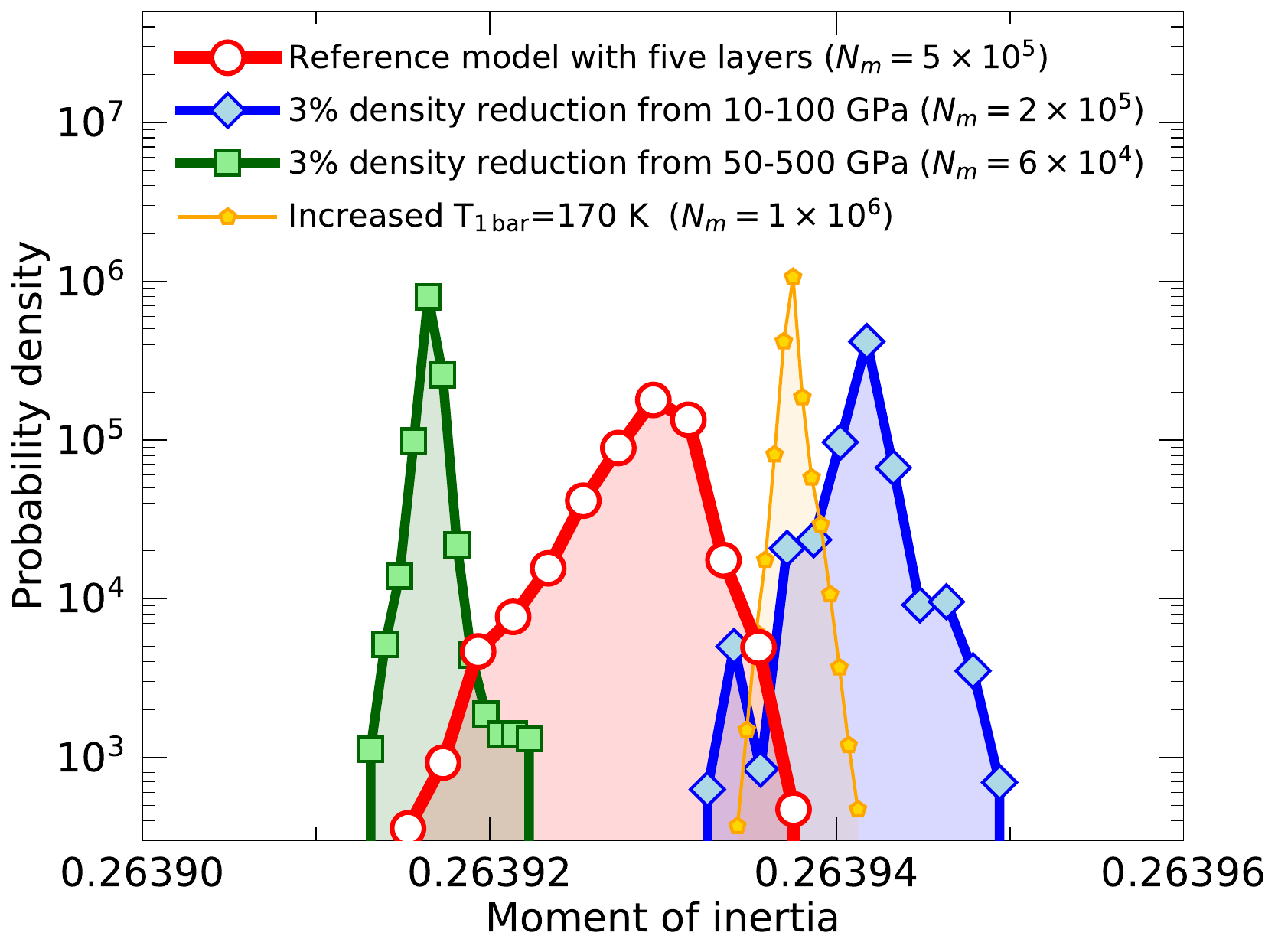}{0.6\textwidth}{}}
\caption{ Probability density distributions of normalized moments of inertia derived from different ensembles of interior models matching even and odd gravity harmonics up to $J_{10}$. All models include a dilute core and contributions from winds. The red circles represent our reference models with five layers. When four layer models are constructed by removing either the helium rain layer or the core transition layer, the MoI distribution hardly changes. If the 1-bar temperature is increased from 166.1 to 170~K, the MoI increases by only $\sim$$10^{-5}$. Slightly larger changes are seen when the density of the hydrogen-helium mixture is reduced over pressure intervals from 10--100 or from 50--500 GPa. In the caption, we specify the size of the ensemble that was used for each histogram.
\label{fig:JupiterMoI}}
\end{figure}

In Fig.~\ref{fig:JupiterMoI}, we compared the MoI from ensembles of interior+wind models that match the entire set of {\em Juno}'s even and odd gravity coefficients up to $J_{10}$~\citep{Durante2020}. The posterior distribution of our five-layer reference ensemble is centered around the MoI value of 0.26393, which we consider to be our most plausible prediction for Jupiter's MoI. If we increased the 1 bar temperature to 170~K, the resulting ensemble of MoI shifted to higher MoI values by a modest amount of $\sim7 \times 10^{-6}$. Slightly larger shifts were obtained when we changed the H-He EOS by reducing the density by 3\% over a selected pressure interval~\citep{DiluteCore2022}. The largest positive shift was obtained for a density reduction from 10--100~GPa and the largest negative was seen if the density was reduced from 50--100~GPa. Both MoI shifts were on the order to $10^{-5}$, which is why we report 0.26393$\pm$0.00001 for Jupiter's MoI.

\begin{figure}
\plotone{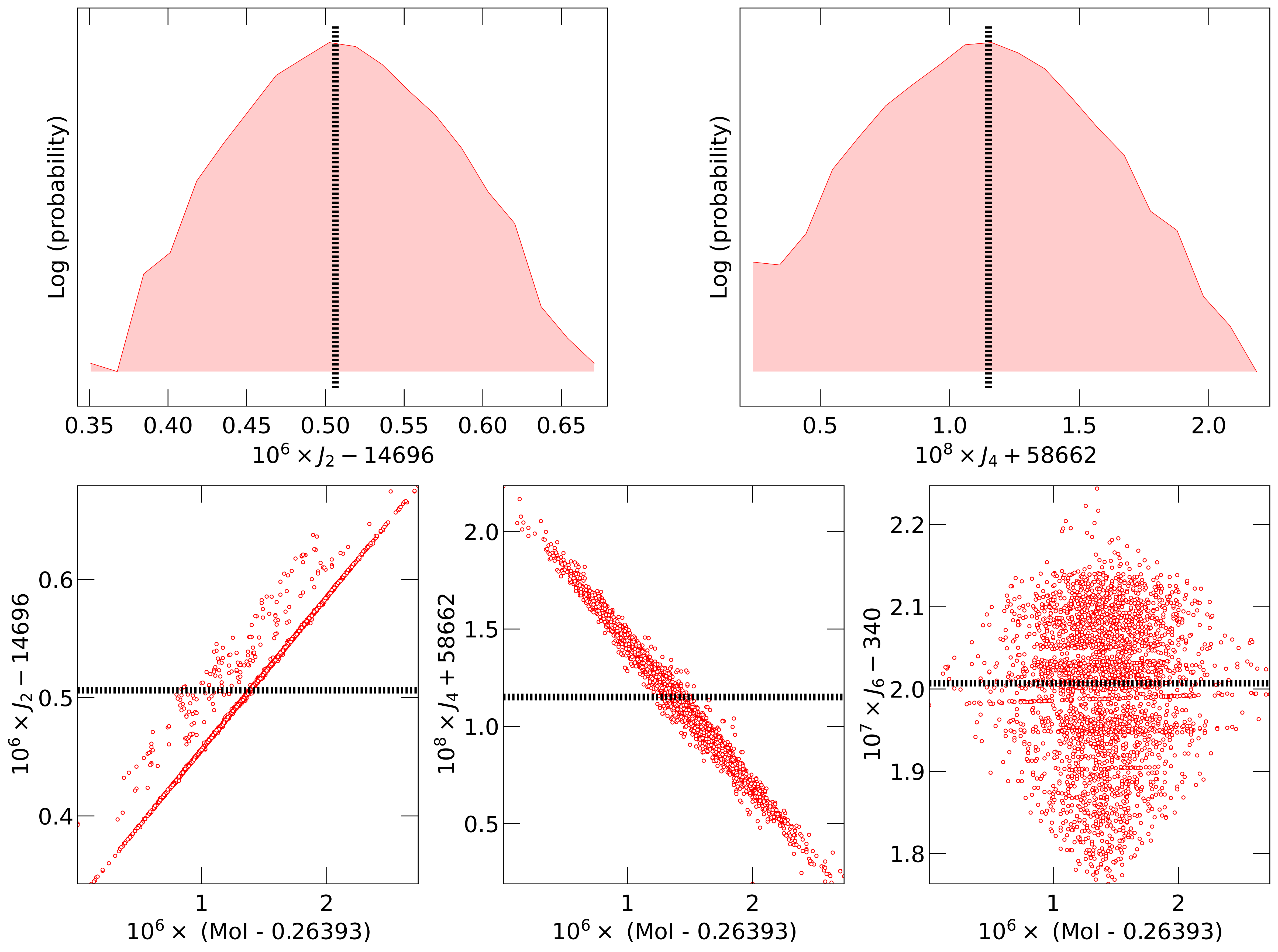}
\caption{ The two upper panels show posterior distributions of gravity harmonics $J_2$ and $J_4$ derived from an ensemble of five layer models. The lower planels illustrate how the computed MoI correlates with $J_2$, $J_4$, and $J_6$. The dashes lines indicate the {\em Juno} gravity measurements from \citet{Durante2020} in Tab.~\ref{tab:JS}.
\label{fig:correlation_plot}}
\end{figure}

In Fig.~\ref{fig:correlation_plot} we plot results from an ensemble of five layer models in order to show how the computed MoI correlates with different gravity harmonics. The MoI correlates positively with $J_2$, negatively with $J_4$, and not in a significant way with $J_6$. (The correlations differ from predictions of two and three layer models in Fig.~\ref{fig:23layers} because they only match the Jupiter's mass and $J_2$.) While the sign and slopes of the correlation of the MoI with $J_2$ and $J_4$ in Fig.~\ref{fig:correlation_plot} differ, one needs to consider that the sign and the magnitude of $J_2$ and $J_4$ differ as well (see Tab.~\ref{tab:JS}). If one removes that dependence by evaluating $J_2 \frac{\partial \rm MoI}{\partial  J_2}=10^{-7}$ and $J_4 \frac{\partial \rm MoI}{\partial  J_4}=8 \times 10^{-8}$, one finds the correlations between the MoI and both gravity coefficients are rather similar. The small magnitudes of $\sim$$10^{-7}$ illustrates that an individual gravity coefficient would need to change a lot to alter the MoI significantly. Fig.~\ref{fig:correlation_plot} also shows that the posterior distributions of $J_2$ and $J_4$ are centered at the {\em Juno} gravity measurements as expected.  

\begin{figure}
\plotone{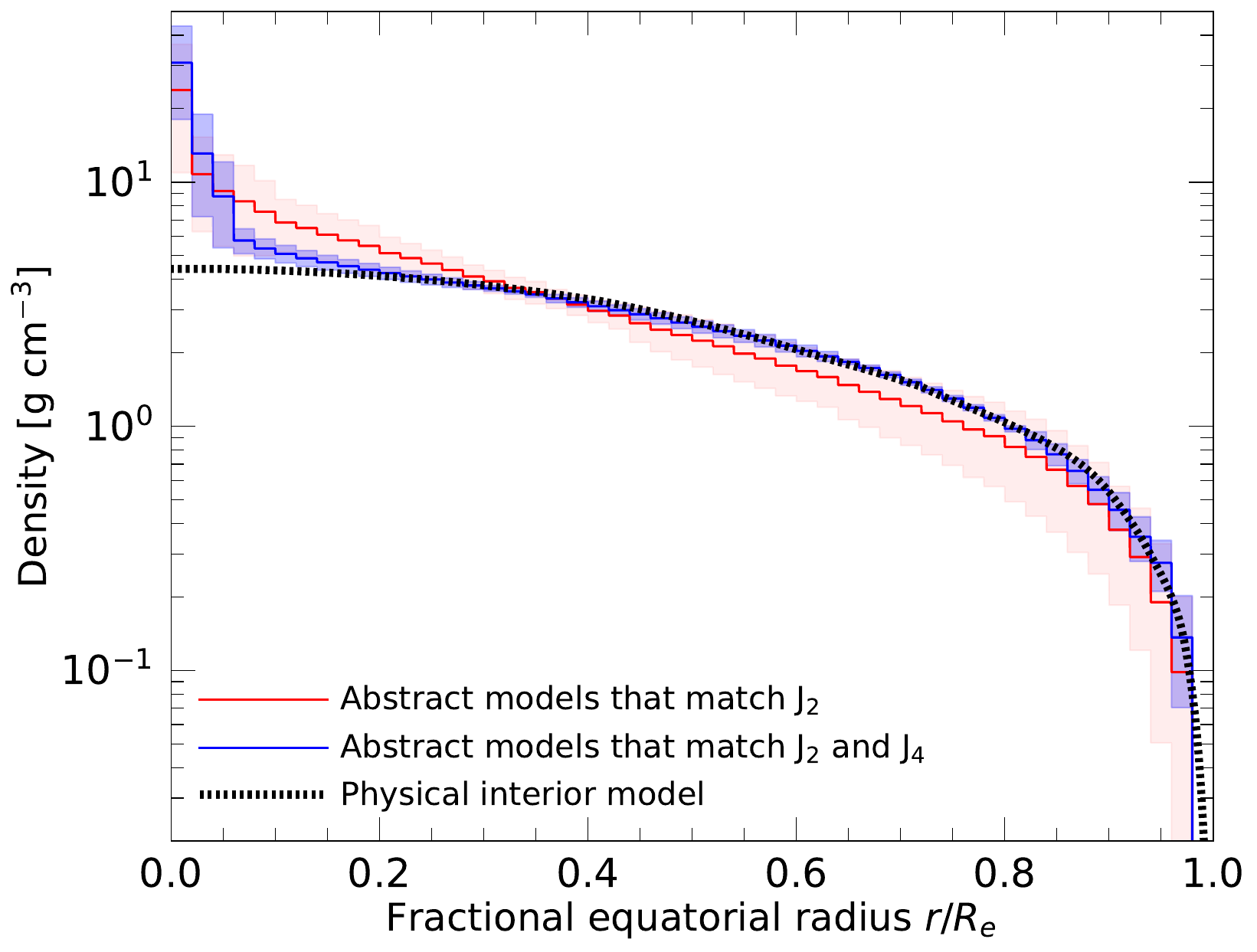}
%\gridline{\fig{density_vs_r_plot01.pdf}{0.6\textwidth}{}}
\caption{ Comparison of the density profiles of abstract models with that of our physical reference model with five layers. The shaded regions represent the standard deviations in density among the abstract models for a given radius. 
\label{fig:density_vs_r_plot}}
\end{figure}

In Fig.~\ref{fig:density_vs_r_plot}, we demonstrate fairly good agreement between the density profiles of our abstract and physical models for Jupiter's interior. For a fractional radius of 0.2 and larger, the density of our physical five layer reference model falls within one standard deviation from the mean of the abstract ensemble that matches the planet's mass, equatorial radius and the gravity coefficients $J_2$ and $J_4$. Both gravity coefficients do not constrain the core region very well and the abstract models can thus yield larger density values there. As expected, models that are only constrained by $J_2$ show a wider range of density values for given radius. Larger density values favored for $r<0.3$ and smaller values for $r>0.4$. Still for most radii, we find that the predictions from the $J_2$ and $J_4$ constrained models fall within one standard deviation of the $J_2$ constrained models.

\begin{table}
\centering                                                        
\begin{tabular}{lll}
\toprule
Jupiter's MoI=$C/(MR_e^2)$& Method and assumptions & Reference \\
%\hline
%0.235 $\to$ 0.264 & MoI change during an assumed path for Jupiter's evolution & \citet{Helled_2012}\\
\hline
0.26401 & Third-order ToF, {\em Pioneer} and {\em Voyager} data & \citet{HubbardMarley1989}\\
0.2640  & Consistent level curve method, {\em Pioneer} and {\em Voyager} data & \citet{Wisdom1996}\\
0.2513~ -- 0.2528$^\ddag$~& ToF, {\em Pioneer} and {\em Voyager} data & \citet{Helled_2011}\\
0.25578 -- 0.27160 & ToF, three layer models, JUP230$^*$ & \citet{Nettelmann_2012}\\
0.26381 -- 0.26399 & CMS, compact core models, DFT and SC EOS, JUP230$^*$ & \citet{HubbardMilitzer2016}\\
0.26391 -- 0.26403 & CMS, dilute and compact core models, physical EOS, {\em Juno} data & \citet{Wahl2017b}\\
0.2629~  -- 0.2641$^\dagger$~ & ToF, empirical EOS, earliest {\em Juno} data & \citet{Ni2018}\\
0.26341 -- 0.26387 & ToF, polytropic and polynomial EOS, {\em Juno} data & \citet{Neuenschwander_2021}\\
\hline
0.26027 -- 0.26477 & Abstract models with 50 spheroids that match only {\em Juno}'s $J_2$ & this work, Fig.~\ref{fig:J2J4J6}\\
0.26385 -- 0.26400 & Physical two and three layer models, CMS, only match {\em Juno's} $J_2$ & this work, Fig.~\ref{fig:23layers}\\
0.26387  --  0.26401 & Abstract models with 50 spheroids that match {\em Juno}'s $J_2$ and $J_4$ & this work, Fig.~\ref{fig:J2J4J6}\\
0.26393 -- 0.26398 & Abstract models with 50 spheroids that match {\em Juno}'s $J_2-J_6$ & this work, Fig.~\ref{fig:J2J4J6}\\
0.26393 $\pm$ 0.00001 & Five layer model, physical EOS, CMS, match all {\em Juno}'s $J_2-J_{10}$ & this work, Fig.~\ref{fig:JupiterMoI}\\
\hline
\end{tabular}
\caption{Predictions for Jupiter's MoI ($R_e=71492$~km) derived under different assumptions.
$^*$JUP230 refers to \citet{jac03}. 
$^\dagger$Converted using mean radius of $R_m=69911$~km (private communication with author.) $^\ddag$Converted using $R_m=69893.175$~km \citep{Helled_2012}.
\label{tab:MoI}}
\end{table}

\begin{table}
\centering                                                        
\begin{tabular}{lcllll}
\toprule
Planet/reference & Rotation period & $J_2 \times 10^6$ & $J_4 \times 10^6$ & $J_6 \times 10^6$ & $C/MR_e^2$\\
\hline
{\em Jupiter} & 9:55:29.711 h\\
Our 5 layer model \citet{DiluteCore2022} & & 14696.5063 & --586.6085 & 34.2007 & 0.26393\\
Three layer model with $T_{\rm 1 bar}=183$ K \citet{Miguel2022}     & & 14698 & --586.6 & 34.11 & 0.26391\\
Two models from \cite{Nettelmann2021}    & & 14719 & --587.7 & 34.30 & 0.26413\\
                                         & & 14723 & --587.7 & 34.24 & 0.26419\\
\hline
{\em Saturn}\\
Our preferred model with DR from \citet{Militzer2019a}$^\dagger$ & 10:33:34 h & 16324.1078 & --939.1687 & 86.8743 & 0.21814 \\
Model from \cite{Nettelmann2021}$^\dagger$       & 10:33:34 h & 16334.2 & --940.149 & 84.208 & 0.21873\\
Two models from \citet{Mankovich2021}            & 10:33:38 h & 16327.4 & --939.507 & 84.686 & 0.21876\\ 
                                                 & 10:33:38 h & 16332.1 & --939.835 & 84.603 & 0.21879\\
\hline
\end{tabular}
\caption{Comparison with Saturn and Jupiter models from other authors. With the CMS method, we calculated the $J_n$ and MoI values for models that were originally constructed with the theory of figures. $^\dagger$The CMS calculations were performed with $R_e$=60367~km and an outer pressure level of 0.1 bar but the results were rescaled to $R_e$=60268~km.
\label{tab:MoI2}}
\end{table}

In Tab.~\ref{tab:MoI}, we compare our result with different predictions for Jupiter's MoI in the literature. Early determinations based on {\em Pioneer} and {\em Voyager} measurements by \citet{HubbardMarley1989} and \citet{Wisdom1996}, who assumed uniform rotation, predicted Jupiter's MoI to be 0.2640, which is very close to the 0.26393 $\pm$ 0.00001 value that we derived when we match the {\em Juno} measurements with models that included DR effects. This now preferred value is also included in the ranges from earlier CMS calculations by \citet{HubbardMilitzer2016} and \citet{Wahl2017b}. With a low-order ToF, \citet{Helled_2011} predicted smaller MoI values. \citet{Nettelmann_2012} predicted a very wide range of MoI values because not all models were constructed to match $J_2$, $J_4$, and $J_6$. \citet{Ni2018} adopted the approach from \citet{AndersonSchubert2007} when he adjusted coefficients of a polynomial function for the density profile in Jupiter's interior in order to match the first gravity measurements of the {\em Juno} spacecraft. With the theory of figures, he obtained a range for Jupiter's MoI that includes our most reliable value. In the lower part of the table, we show how the range of predicted MoI shrinks when more and more of {\em Juno's} gravity harmonics are reproduced. 

In Tab.~\ref{tab:MoI2}, we compare our predictions for Jupiter and Saturn with results from CMS calculations that we performed for models that other authors had constructed with the theory of figures. The central quantity of this approach is the volumetric radius, $s$, of different interior layers. When we read in the model files from other authors, we construct a density function, $\rho(s)$, that we can interpolate. As our CMS calculation converges step by step towards a self-consistent solution, we calculate volumetric radius of every spheroid and obtain the corresponding density values by interpolation. We then update the density of every layer by averaging the density values of corresponding inner and outer spheroids. After all layer densities have been updated, we scale all densities again to match the total planet mass exactly. We increased the number of spheroids in our CMS calculation up to 65536 to obtain converged results. We found this to be a robust approach to import model files from other authors.

The agreement among the resulting MoI values in Tab.~\ref{tab:MoI2} is very good even though some residual differences can be expected because the theory of figures is a perturbative approach that neglects high order terms. Furthermore not every model was constructed to match the measured gravity field with the same level of precision. Finally planetary interior models are complex and authors invoke an array of not always compatible set of assumptions. For example, while we invoke the concept of a dilute core and combine it with a model for the planet's winds to match {\em Juno's} $J_4$ and $J_6$ measurements, \citet{Miguel2022} succeeded in doing so by raising the 1~bar temperature from 166 to 183~K when ensembles of traditional three layer models were constructed. Still the gravity coefficients and computed MoI are in good agreement with those of our five layer model. The MoI values, that we computed for two models by \citet{Nettelmann2021}, were $2 \times 10^{-4}$ larger than our predictions. We attributed this difference to the fact that we obtained with our CMS calculations a $J_2$ value that was $2 \times 10^{-5}$ higher than the {\em Juno} measurements. 

In Tab.~\ref{tab:MoI2}, we also compare the predictions of four Saturn models that were constructed for a rotation period of 10:33:34~h that \citet{Militzer2019a} derived by matching the planet's polar radius or for a very similar period of 10:33:38~h that \citet{Mankovich2019} derived from ring-seismological calculations. The CMS calculations for models by \citet{Mankovich2021} and \citet{Nettelmann2021} yielded MoI values that were $\sim 6 \times 10^{-4}$ larger than that of our preferred Saturn model with DR. We primarily attribute this modest difference to the fact \citet{Mankovich2021} and \citet{Nettelmann2021} do not have DR in their models and thus make no attempt to match the observed $J_6$ value. Overall the results in Tab.~\ref{tab:MoI2} confirm that a planet's MoI is very well constrained by measurements of the gravity coefficients $J_2$, $J_4$, and $J_6$.

%%%%%%%%%%%%%%%%%%%%%%%%%%%%%%%%%%%%%%%%%%%%%%%%%%%%%%%%%%%%%%%%%%%%%%%%%%%%%%%%%%%%%%%%
%%%%%%%%%%%%%%%%%%%%%%%%%%%%%%%%%%%%%%%%%%%%%%%%%%%%%%%%%%%%%%%%%%%%%%%%%%%%%%%%%%%%%%%%
%%%%%%%%%%%%%%%%%%%%%%%%%%%%%%%%%%%%%%%%%%%%%%%%%%%%%%%%%%%%%%%%%%%%%%%%%%%%%%%%%%%%%%%%

\section{Conclusions}
\label{conclusions}

With nonperturbative concentric Maclaurin spheroid method, we construct models for the interiors of Jupiter and Saturn under a number of different assumption. Our ensemble includes physical models based on a realistic EOS for hydrogen and helium, and abstract models with a small number of constant density spheroids. For both sets of assumptions we find that current spacecraft measurements of the Jupiter and Saturn gravity fields constrain the planets' moment of inertia (MoI) fairly tightly, but then zonal winds (or differential rotation, DR) emerge as the leading source of MoI uncertainty, assuming the planets' rotation rates have been constrained (by magnetic field measurements for Jupiter or by observations of the polar radius for Saturn.)

If DR effects are excluded, the gravity coefficients $J_2$, $J_4$, and $J_6$ one-by-one constrain the predicted MoI more and more tightly. Already mass, equatorial radius and $J_2$ alone constrain Saturn's MoI by $\sim$10\% while Jupiter's MoI is constrained to a level of $\sim$1\%. If models are required to match also $J_4$, the range of Saturn's and Jupiter's MoI shrinks to 3\% and 0.05\%. If models match also $J_6$, the allowed MoI range shrinks to 0.07\% and 0.008\%, respectively.

%subject to two effects that can dominate the uncertainties. The inferred MoI of both planets are affected by zonal winds (or differential rotation). For Saturn, uncertainty in the planet's rotation period introduces a further 2\% uncertainty in the planet's MoI. If the rotation period is constrained by observations of the planet's polar radius, the uncertainty of the MoI reduces to $\sim$0.2\%.

%As expected, a giant planet's MoI is determined almost exclusively by its interior structure while DR effects make only negligible contributions directly. 
However, DR effects can make significant contributions to the gravity harmonics $J_6$ and thereby alter the $J_6$ term that needs to come from the interior structure if interior+wind models are constructed to match specific spacecraft measurements. We find that Saturn's MoI drops by 0.4\% when effects of DR are added to interior models that match the gravity harmonics $J_2$, $J_4$, and $J_6$. In principle, such a drop could be detected by a direct precise MoI measurement by a spacecraft that orbits Saturn over a sufficiently long arc of Saturn's precession. 

This 0.4\% drop of Saturn's MoI is mainly caused by the way models match the gravity coefficient $J_6$. On Saturn the zonal winds are predicted to reach a depth of $\sim$9000~km \citep{Iess2018} and involve 7\% of the planet's mass. The DR contributions to $J_6$ were thus found to be rather large, on the order of 6\%. For Jupiter, the winds reach only $\sim$3000~km deep \citep{Kaspi2018} and involve only 1\% of the planet's mass. So we estimate the contributions from DR to $J_6$ to be only on the order of 0.8\%. DR effects thus lower Jupiter's MoI by only 0.01\%, too small to be detected by the {\em Juno} spacecraft. %\citet{LaMaistre2016} estimated that the {\em Juno} spacecraft will determine Jupiter's MoI to 0.1\% by measuring the planet's precession rate.

%Uncertainties of Saturn's MoI are dominated by DR effects and uncertainties in the planet's rotation period. Jupiter's MoI is much less affected by DR effects. Therefore they contribute to the uncertainties of the planet's MoI at the same level as other modeling assumptions for the plant's interior structure.

Our models with DR predict Saturn's MoI to be 0.2181$\pm$0.0002. This is 1\% too small for Saturn to be in a spin-orbit resonance with Neptune today but \citet{Wisdom2022} predicted the planet was in resonance in the past when it had an additional moon that was tidally disrupted and formed the rings. With physical but simplified models for Jupiter's interior that match only $J_2$, we obtain wide range from 0.26385–0.26400 for the planet's MoI. For our abstract models with 50 spheroids for Jupiter's interior that match the measured harmonics $J_2$, $J_4$ and $J_6$, we derived a narrower range of possible MoI values from 0.26393-0.26398. Finally with our most plausible five layer models for Jupiter's interior, we predict the planet's MoI to be 0.26393 $\pm$ 0.00001, which is about $\sim$10\% above the critical value of $C/MR^2=0.236$ for the planet to be in spin-orbit resonance with Uranus today~\citep{Ward_Canup_2006}.
% I assume Ward used Jupiter's equatorial but not its volumetric radius since he also use equations with J2

\citet{Wisdom2022} argue that available high-precision measurements of Saturn's zonal harmonics suffice to infer a tight MoI range that rules out a current Saturn precession resonance with Neptune.  By the same token, our predicted range for Jupiter's MoI needs to lie within the range constrained by {\em Juno}'s extended mission measurement of MoI.

\section*{Acknowledgments} 

We thank C. Mankovich, N. Nettelmann, and T. Guillot for sharing model files. The work was funded by the NASA mission Juno. BM also received support from the Center for Matter at Atomic Pressures, which is funded by the U.S. National Science Foundation (PHY-2020249).

\bibliography{jupiter}{}

\begin{thebibliography}{}
\expandafter\ifx\csname natexlab\endcsname\relax\def\natexlab#1{#1}\fi
\providecommand{\url}[1]{\href{#1}{#1}}
\providecommand{\dodoi}[1]{doi:~\href{http://doi.org/#1}{\nolinkurl{#1}}}
\providecommand{\doeprint}[1]{\href{http://ascl.net/#1}{\nolinkurl{http://ascl.net/#1}}}
\providecommand{\doarXiv}[1]{\href{https://arxiv.org/abs/#1}{\nolinkurl{https://arxiv.org/abs/#1}}}

\bibitem[{Anderson \& Schubert(2007)}]{AndersonSchubert2007}
Anderson, J.~D., \& Schubert, G. 2007, Science, 317, 1384,
  \dodoi{10.1126/science.1144835}

\bibitem[{Bolton {et~al.}(2017)Bolton, Adriani, Adumitroaie, Allison, Anderson,
  Atreya, Bloxham, Brown, Connerney, DeJong, Folkner, Gautier, Grassi, Gulkis,
  Guillot, Hansen, Hubbard, Iess, Ingersoll, Janssen, Jorgensen, Kaspi, Levin,
  Li, Lunine, Miguel, Mura, Orton, Owen, Ravine, Smith, Steffes, Stone,
  Stevenson, Thorne, Waite, Durante, Ebert, Greathouse, Hue, Parisi, Szalay, \&
  Wilson}]{Bolton2017}
Bolton, S.~J., Adriani, A., Adumitroaie, V., {et~al.} 2017, Science, 356, 821,
  \dodoi{10.1126/science.aal2108}

\bibitem[{Bourda \& Capitaine(2004)}]{RadauDarwin}
Bourda, G., \& Capitaine, N. 2004, A\&A, 428, 691,
  \dodoi{10.1051/0004-6361:20041533}

\bibitem[{Brygoo {et~al.}(2021)Brygoo, Loubeyre, Millot, Rygg, Celliers,
  Eggert, Jeanloz, \& Collins}]{Brygoo2021}
Brygoo, S., Loubeyre, P., Millot, M., {et~al.} 2021, Nature, 593, 517,
  \dodoi{10.1038/s41586-021-03516-0}

\bibitem[{Campbell \& Anderson(1989)}]{CA89}
Campbell, J.~K., \& Anderson, J.~D. 1989, Astron. J., 97, 1485

\bibitem[{Campbell \& Synnott(1985)}]{CS85}
Campbell, J.~K., \& Synnott, S.~P. 1985, Astron. J., 90, 364

\bibitem[{Cao \& Stevenson(2017)}]{Cao2017}
Cao, H., \& Stevenson, D.~J. 2017, J. Geophys. Res. Planets, 122, 686,
  \dodoi{10.1002/2017JE005272}

\bibitem[{Celliers {et~al.}(2010)Celliers, Loubeyre, Eggert, Brygoo,
  McWilliams, Hicks, Boehly, Jeanloz, \& Collins}]{Celliers2010}
Celliers, P.~M., Loubeyre, P., Eggert, J.~H., {et~al.} 2010, Phys. Rev. Lett.,
  104, 184503

\bibitem[{Collins {et~al.}(1998)Collins, Silva, Celliers, Gold, Foord, Wallace,
  Ng, Weber, Budil, \& Cauble}]{Co98}
Collins, G.~W., Silva, L. B.~D., Celliers, P., {et~al.} 1998, Science, {281},
  1178

\bibitem[{{Da~Silva} {et~al.}(1997){Da~Silva}, Celliers, Collins, Budil,
  Holmes, Barbee, Hammel, Kilkenny, Wallace, Ross, Cauble, Ng, \& Chiu}]{Si97}
{Da~Silva}, I.~B., Celliers, P., Collins, G.~W., {et~al.} 1997, Phys. Rev.
  Lett., {78}, 483

\bibitem[{Debras \& Chabrier(2019)}]{Debras2019}
Debras, F., \& Chabrier, G. 2019, The Astrophysical Journal, 872, 100,
  \dodoi{10.3847/1538-4357/aaff65}

\bibitem[{Debras {et~al.}(2021)Debras, Chabrier, \& Stevenson}]{Debras2021}
Debras, J., Chabrier, G., \& Stevenson, D.~J. 2021, Astrop. J. Lett., 913, 21

\bibitem[{Dietrich {et~al.}(2021)Dietrich, Wulff, Wicht, \&
  Christensen}]{Dietrich2021}
Dietrich, W., Wulff, P., Wicht, J., \& Christensen, U.~R. 2021, Monthly Notices
  of the Royal Astronomical Society, 505, 3177, \dodoi{10.1093/mnras/stab1566}

\bibitem[{Durante {et~al.}(2020)Durante, Buccino, Tommei, Parisi, Serra,
  Zannoni, Notaro, Racioppa, Lari, Casajus, Iess, Folkner, Tortora, \&
  Bolton}]{Durante2020}
Durante, D., Buccino, D.~R., Tommei, G., {et~al.} 2020, Geophys. Res. Lett.,
  47, e2019GL086572

\bibitem[{Fortney {et~al.}(2016)Fortney, Helled, Nettelmann, Stevenson, Marley,
  Hubbard, \& Iess}]{Fortney2016}
Fortney, J.~J., Helled, R., Nettelmann, N., {et~al.} 2016, in Saturn in the
  21st Century, ed. B.~Kevin, M.~Flasar, N.~Krupp, \& S.~Thomas (Cambridge
  University Press), 1--28.
\newblock \doarXiv{1609.06324}

\bibitem[{French {et~al.}(2009)French, Mattsson, Nettelmann, \&
  Redmer}]{french-prb-09}
French, M., Mattsson, T.~R., Nettelmann, N., \& Redmer, R. 2009, Phys. Rev. B,
  79, 054107

\bibitem[{Fuller {et~al.}(2016)Fuller, Luan, \& Quataert}]{Fuller_2016}
Fuller, J., Luan, J., \& Quataert, E. 2016, Monthly Notices of the Royal
  Astronomical Society, 458, 3867, \dodoi{10.1093/mnras/stw609}

\bibitem[{Galanti \& Kaspi(2020)}]{Galanti2021}
Galanti, E., \& Kaspi, Y. 2020, Monthly Notices of the Royal Astronomical
  Society, 501, 2352, \dodoi{10.1093/mnras/staa3722}

\bibitem[{Garc{\'{i}}a-Melendo {et~al.}(2011)Garc{\'{i}}a-Melendo,
  P{\'{e}}rez-Hoyos, S{\'{a}}nchez-Lavega, \& Hueso}]{GM2011}
Garc{\'{i}}a-Melendo, E., P{\'{e}}rez-Hoyos, S., S{\'{a}}nchez-Lavega, A., \&
  Hueso, R. 2011, Icarus, 215, 62, \dodoi{10.1016/j.icarus.2011.07.005}

\bibitem[{Guillot(2005)}]{Guillot2005}
Guillot, T. 2005, Ann. Rev. Earth Planet. Sci., 33, 493

\bibitem[{Guillot {et~al.}(2004)Guillot, Stevenson, Hubbard, \&
  Saumon}]{Guillot2004}
Guillot, T., Stevenson, D.~J., Hubbard, W.~B., \& Saumon, D. 2004, In: Jupiter.
  The planet, 35

\bibitem[{Gupta {et~al.}(2022)Gupta, Atreya, Steffes, Fletcher, Guillot,
  Allison, Bolton, Helled, Levin, Li, Lunine, Miguel, Orton, Waite, \&
  Withers}]{Gupta_2022}
Gupta, P., Atreya, S.~K., Steffes, P.~G., {et~al.} 2022, The Planetary Science
  Journal, 3, 159, \dodoi{10.3847/psj/ac6956}

\bibitem[{Helled(2012)}]{Helled_2012}
Helled, R. 2012, The Astrophysical Journal, 748, L16,
  \dodoi{10.1088/2041-8205/748/1/l16}

\bibitem[{Helled {et~al.}(2011)Helled, Anderson, Schubert, \&
  Stevenson}]{Helled_2011}
Helled, R., Anderson, J.~D., Schubert, G., \& Stevenson, D.~J. 2011, Icarus,
  216, 440, \dodoi{https://doi.org/10.1016/j.icarus.2011.09.016}

\bibitem[{Helled {et~al.}(2009)Helled, Schubert, \& Anderson}]{Helled_2009}
Helled, R., Schubert, G., \& Anderson, J.~D. 2009, Icarus, 199, 368,
  \dodoi{https://doi.org/10.1016/j.icarus.2008.10.005}

\bibitem[{Hubbard \& Marley(1989)}]{HubbardMarley1989}
Hubbard, W., \& Marley, M.~S. 1989, Icarus, 78, 102,
  \dodoi{https://doi.org/10.1016/0019-1035(89)90072-9}

\bibitem[{Hubbard(2013)}]{Hubbard2013}
Hubbard, W.~B. 2013, Astrophys. J., 768, 43, \dodoi{10.1088/0004-637X/768/1/43}

\bibitem[{Hubbard \& Militzer(2016{\natexlab{a}})}]{HM16}
Hubbard, W.~B., \& Militzer, B. 2016{\natexlab{a}}, Astrophys. J., 820, 80

\bibitem[{Hubbard \& Militzer(2016{\natexlab{b}})}]{HubbardMilitzer2016}
---. 2016{\natexlab{b}}, Astrophys. J., 820, 80

\bibitem[{Iess {et~al.}(2018)Iess, Folkner, Durante, Parisi, Kaspi, Galanti,
  Guillot, Hubbard, Stevenson, Anderson, Buccino, Casajus, Milani, Park,
  Racioppa, Serra, Tortora, Zannoni, Cao, Helled, Lunine, Miguel, Militzer,
  Wahl, Connerney, Levin, \& Bolton}]{Iess2018}
Iess, L., Folkner, W., Durante, D., {et~al.} 2018, Nature, 555,
  \dodoi{10.1038/nature25776}

\bibitem[{Iess {et~al.}(2019)Iess, Militzer, Kaspi, Nicholson, Durante,
  Racioppa, Anabtawi, Galanti, Hubbard, Mariani, Tortora, Wahl, \&
  Zannoni}]{Iess2019}
Iess, L., Militzer, B., Kaspi, Y., {et~al.} 2019, Science, 2965, eaat2965,
  \dodoi{10.1126/science.aat2965}

\bibitem[{Jacobson(2003)}]{jac03}
Jacobson, R.~A. 2003, https://ssd.jpl.nasa.gov/tools/gravity.html, Outer
  planets, JUP230 orbit solution

\bibitem[{Kaspi {et~al.}(2016)Kaspi, Davighi, Galanti, \& Hubbard}]{Kaspi2016}
Kaspi, Y., Davighi, J.~E., Galanti, E., \& Hubbard, W.~B. 2016, Icarus, 276,
  170

\bibitem[{Kaspi {et~al.}(2018)Kaspi, Galanti, Hubbard, Stevenson, Bolton, Iess,
  Guillot, Bloxham, Connerney, Cao, Durante, Folkner, Helled, Ingersoll, Levin,
  Lunine, Miguel, Militzer, Parisi, \& Wahl}]{Kaspi2018}
Kaspi, Y., Galanti, E., Hubbard, W., {et~al.} 2018, Nature, 555,
  \dodoi{10.1038/nature25793}

\bibitem[{Knudson {et~al.}(2001)Knudson, Hanson, Bailey, Hall, Asay, \&
  Anderson}]{Kn01}
Knudson, M.~D., Hanson, D.~L., Bailey, J.~E., {et~al.} 2001, Phys. Rev. Lett.,
  87, 225501

\bibitem[{Ledoux(1947)}]{Ledoux1947}
Ledoux, P. 1947, Astrophys. J. Lett., 105, 305

\bibitem[{Lindal {et~al.}(1981)Lindal, Wood, Levy, Anderson, Sweetnam, Hotz,
  Buckles, Holmes, Doms, Eshleman, Tyler, \& Croft}]{Lindal1981}
Lindal, G.~F., Wood, G.~E., Levy, G.~S., {et~al.} 1981, Journal of Geophysical
  Research: Space Physics, 86, 8721, \dodoi{10.1029/JA086iA10p08721}

\bibitem[{Lodders(2010)}]{Lodders2010}
Lodders, K. 2010, in Astrophysics and Space Science Proceedings, ed. A.~Goswami
  \& B.~E. Reddy (Berlin: Springer-Verlag), 379--417

\bibitem[{Mankovich {et~al.}(2019)Mankovich, Marley, Fortney, \&
  Movshovitz}]{Mankovich2019}
Mankovich, C., Marley, M.~S., Fortney, J.~J., \& Movshovitz, N. 2019, The
  Astrophysical Journal, 871, 1, \dodoi{10.3847/1538-4357/aaf798}

\bibitem[{Mankovich \& Fuller(2021)}]{Mankovich2021}
Mankovich, C.~R., \& Fuller, J. 2021, Nature Astronomy, 5, 1103,
  \dodoi{10.1038/s41550-021-01448-3}

\bibitem[{Miguel {et~al.}(2022)Miguel, Bazot, Guillot, Howard, Galanti, Kaspi,
  Hubbard, Militzer, Helled, Atreya, Connerney, Durante, Kulowski, Lunine,
  Stevenson, \& Bolton}]{Miguel2022}
Miguel, Y., Bazot, M., Guillot, T., {et~al.} 2022, Astron. and Astrophys., 662,
  A18

\bibitem[{Militzer(2006)}]{Mi06}
Militzer, B. 2006, Phys. Rev. Lett., 97, 175501

\bibitem[{Militzer(2009)}]{Mi09}
---. 2009, Phys. Rev. B, 79, 155105

\bibitem[{Militzer(2013)}]{Militzer2013}
---. 2013, Phys. Rev. B, 87, 014202

\bibitem[{Militzer(2023)}]{Militzer_QMC_2023}
---. 2023, The Astrophysical Journal, 953, 111,
  \dodoi{10.3847/1538-4357/ace1f1}

\bibitem[{Militzer \& Hubbard(2013)}]{MH13}
Militzer, B., \& Hubbard, W.~B. 2013, Astrophys. J., 774, 148

\bibitem[{Militzer {et~al.}(2008)Militzer, Hubbard, Vorberger, Tamblyn, \&
  Bonev}]{MHVTB}
Militzer, B., Hubbard, W.~H., Vorberger, J., Tamblyn, I., \& Bonev, S.~A. 2008,
  Astrophys. J. Lett., 688, L45

\bibitem[{Militzer {et~al.}(2016)Militzer, Soubiran, Wahl, \&
  Hubbard}]{MilitzerJGR2016}
Militzer, B., Soubiran, F., Wahl, S.~M., \& Hubbard, W. 2016, J. Geophys. Res.
  Planets, 121, 1552

\bibitem[{Militzer {et~al.}(2019)Militzer, Wahl, \& Hubbard}]{Militzer2019a}
Militzer, B., Wahl, S., \& Hubbard, W.~B. 2019, The Astrophysical Journal, 879,
  78, \dodoi{10.3847/1538-4357/ab23f0}

\bibitem[{Militzer {et~al.}(2022)Militzer, Hubbard, Wahl, Lunine, Galanti,
  Kaspi, Miguel, Guillot, Moore, Parisi, Connerney, Helled, Cao, Mankovich,
  Stevenson, Park, Wong, treya, Anderson, \& Bolton}]{DiluteCore2022}
Militzer, B., Hubbard, W.~B., Wahl, S., {et~al.} 2022, Planet. Sci. J., 3, 185

\bibitem[{Millot {et~al.}(2020)Millot, Zhang, Fratanduono, Coppari, Hamel,
  Militzer, Simonova, Shcheka, Dubrovinskaia, Dubrovinsky, \&
  Eggert}]{Millot2020}
Millot, M., Zhang, S., Fratanduono, D.~E., {et~al.} 2020, Geophysical Research
  Letters, 47, e2019GL085476, \dodoi{https://doi.org/10.1029/2019GL085476}

\bibitem[{Morales {et~al.}(2013)Morales, McMahon, Pierleonie, \&
  Ceperley}]{Morales2013}
Morales, M.~A., McMahon, J.~M., Pierleonie, C., \& Ceperley, D.~M. 2013, Phys.
  Rev. Lett., 110, 065702

\bibitem[{Morales {et~al.}(2009)Morales, Pierleoni, Schwegler, \&
  Ceperley}]{Morales2009}
Morales, M.~A., Pierleoni, C., Schwegler, E., \& Ceperley, D.~M. 2009, Proc.
  Nat. Acad. Sci., 106, 1324

\bibitem[{Morales {et~al.}(2010)Morales, Pierleoni, Schwegler, \&
  Ceperley}]{Morales2010}
---. 2010, Proc. Nat. Acad. Sci., 107, 12799

\bibitem[{Movshovitz {et~al.}(2020)Movshovitz, Fortney, Mankovich, Thorngren,
  \& Helled}]{Movshovitz_2020}
Movshovitz, N., Fortney, J.~J., Mankovich, C., Thorngren, D., \& Helled, R.
  2020, The Astrophysical Journal, 891, 109, \dodoi{10.3847/1538-4357/ab71ff}

\bibitem[{Nettelmann(2017)}]{Nettelmann2017}
Nettelmann, N. 2017, Astronomy {\&} Astrophysics, 606, A139,
  \dodoi{10.1051/0004-6361/201731550}

\bibitem[{Nettelmann {et~al.}(2012{\natexlab{a}})Nettelmann, Becker, Holst, \&
  Redmer}]{Nettelmann_2012}
Nettelmann, N., Becker, A., Holst, B., \& Redmer, R. 2012{\natexlab{a}}, The
  Astrophysical Journal, 750, 52, \dodoi{10.1088/0004-637x/750/1/52}

\bibitem[{Nettelmann {et~al.}(2012{\natexlab{b}})Nettelmann, Becker, Holst, \&
  Redmer}]{Nettelmann2012}
---. 2012{\natexlab{b}}, Astrophys. J., 750, 52

\bibitem[{Nettelmann {et~al.}(2008)Nettelmann, Holst, Kietzmann, French,
  Redmer, \& Blaschke}]{NHKFRB}
Nettelmann, N., Holst, B., Kietzmann, A., {et~al.} 2008, Astrophys. J., 683,
  1217

\bibitem[{Nettelmann {et~al.}(2013)Nettelmann, P{\"{u}}stow, \& Redmer}]{N13}
Nettelmann, N., P{\"{u}}stow, R., \& Redmer, R. 2013, Icarus, 225, 548,
  \dodoi{10.1016/j.icarus.2013.04.018}

\bibitem[{Nettelmann {et~al.}(2021)Nettelmann, Movshovitz, Ni, Fortney,
  Galanti, Kaspi, Helled, Mankovich, \& Bolton}]{Nettelmann2021}
Nettelmann, N., Movshovitz, N., Ni, D., {et~al.} 2021, Planetary Science
  Journal, 2, 241, \dodoi{10.3847/psj/ac390a}

\bibitem[{Neuenschwander {et~al.}(2021)Neuenschwander, Helled, Movshovitz, \&
  Fortney}]{Neuenschwander_2021}
Neuenschwander, B.~A., Helled, R., Movshovitz, N., \& Fortney, J.~J. 2021, The
  Astrophysical Journal, 910, 38, \dodoi{10.3847/1538-4357/abdfd4}

\bibitem[{Ni(2018)}]{Ni2018}
Ni, D. 2018, A\&A, 613, A32, \dodoi{10.1051/0004-6361/201732183}

\bibitem[{Roulston \& Stevenson(1995)}]{Roulston1995}
Roulston, M., \& Stevenson, D. 1995, EOS, 76, 59–72

\bibitem[{Saillenfest {et~al.}(2021)Saillenfest, Lari, \&
  Bou{\'e}}]{Saillenfest_2021}
Saillenfest, M., Lari, G., \& Bou{\'e}, G. 2021, Nature Astronomy, 5, 345,
  \dodoi{10.1038/s41550-020-01284-x}

\bibitem[{Saillenfest {et~al.}(2022)Saillenfest, Rogoszinski, Lari, Baillié,
  Boué, Crida, \& Lainey}]{Saillenfest_Uranus_2022}
Saillenfest, M., Rogoszinski, Z., Lari, G., {et~al.} 2022,
  https://doi.org/10.48550/arXiv.2209.10590

\bibitem[{Sanchez-Lavega {et~al.}(2000)Sanchez-Lavega, Rojas, \& Sada}]{SL2000}
Sanchez-Lavega, A., Rojas, J.~F., \& Sada, P.~V. 2000, Icarus, 147, 405,
  \dodoi{10.1006/icar.2000.6449}

\bibitem[{Saumon {et~al.}(1995{\natexlab{a}})Saumon, Chabrier, \& Horn}]{SC95}
Saumon, D., Chabrier, G., \& Horn, H. M.~V. 1995{\natexlab{a}}, Astrophys. J.
  Suppl., 99, 713

\bibitem[{Saumon {et~al.}(1995{\natexlab{b}})Saumon, Chabrier, \& van
  Horn}]{Saumon1995}
Saumon, D., Chabrier, G., \& van Horn, H.~M. 1995{\natexlab{b}}, ApJSS, 99, 713

\bibitem[{Saumon \& Guillot(2004{\natexlab{a}})}]{SG04}
Saumon, D., \& Guillot, T. 2004{\natexlab{a}}, Astrophys. J., 609, 1170,
  \dodoi{10.1086/421257}

\bibitem[{Saumon \& Guillot(2004{\natexlab{b}})}]{saumon-apj-04}
---. 2004{\natexlab{b}}, Astrop. J, 609, 1170

\bibitem[{Seiff {et~al.}(1997)Seiff, Kirk, Knight, Young, Milos, Venkatapathy,
  Mihalov, Blanchard, Young, \& Schubert}]{Seiff1998}
Seiff, A., Kirk, D.~B., Knight, T. C.~D., {et~al.} 1997, Science, 276, 102,
  \dodoi{10.1126/science.276.5309.102}

\bibitem[{Soubiran \& Militzer(2015)}]{SoubiranMilitzer2015}
Soubiran, F., \& Militzer, B. 2015, Astrophys. J, 806, 228

\bibitem[{Spilker(2019)}]{Spilker2019}
Spilker, T. 2019, Science, 364, 1046

\bibitem[{Stevenson \& Salpeter(1977)}]{stevenson-astropj-77-i}
Stevenson, D.~J., \& Salpeter, E.~E. 1977, Astrophys J Suppl., 35, 221

\bibitem[{Tollefson {et~al.}(2017)Tollefson, Wong, de~Pater, Simon, Orton,
  Rogers, Atreya, Cosentino, Januszewski, Morales-Juberías, \&
  Marcus}]{Tollefson2017}
Tollefson, J., Wong, M.~H., de~Pater, I., {et~al.} 2017, Icarus, 296, 163,
  \dodoi{https://doi.org/10.1016/j.icarus.2017.06.007}

\bibitem[{von Zahn {et~al.}(1998)von Zahn, Hunten, \&
  Lehmacher}]{vonzahn-jgr-98}
von Zahn, U., Hunten, D.~M., \& Lehmacher, G. 1998, J. Geophys. Res., 103,
  22815

\bibitem[{Vorberger {et~al.}(2007)Vorberger, Tamblyn, Militzer, \&
  Bonev}]{Vo07}
Vorberger, J., Tamblyn, I., Militzer, B., \& Bonev, S. 2007, Phys. Rev. B, 75,
  024206

\bibitem[{Wahl {et~al.}(2017{\natexlab{a}})Wahl, Hubbard, \&
  Militzer}]{Wahl2017b}
Wahl, S.~M., Hubbard, W.~B., \& Militzer, B. 2017{\natexlab{a}}, Icarus, 282,
  183, \dodoi{10.1016/j.icarus.2016.09.011}

\bibitem[{Wahl {et~al.}(2021)Wahl, Thorngren, Lu, \& Militzer}]{Wahl2021}
Wahl, S.~M., Thorngren, D., Lu, T., \& Militzer, B. 2021, Astrophys. J., 921,
  105

\bibitem[{Wahl {et~al.}(2017{\natexlab{b}})Wahl, Hubbard, Militzer, Guillot,
  Miguel, Movshovitz, Kaspi, Helled, Reese, Galanti, Levin, Connerney, \&
  Bolton}]{Wahl2017a}
Wahl, S.~M., Hubbard, W.~B., Militzer, B., {et~al.} 2017{\natexlab{b}},
  Geophys. Res. Lett., 44, 4649, \dodoi{10.1002/2017GL073160}

\bibitem[{Ward \& Canup(2006)}]{Ward_Canup_2006}
Ward, W.~R., \& Canup, R.~M. 2006, The Astrophysical Journal, 640, L91,
  \dodoi{10.1086/503156}

\bibitem[{Wilson \& Militzer(2010)}]{Wilson2010}
Wilson, H.~F., \& Militzer, B. 2010, Phys. Rev. Lett., 104, 121101.
\newblock \url{http://www.ncbi.nlm.nih.gov/pubmed/20366523}

\bibitem[{Wilson \& Militzer(2012{\natexlab{a}})}]{WilsonMilitzer2012b}
---. 2012{\natexlab{a}}, Phys. Rev. Lett., 108, 111101

\bibitem[{Wilson \& Militzer(2012{\natexlab{b}})}]{WilsonMilitzer2012}
---. 2012{\natexlab{b}}, Astrophys. J., 745, 54

\bibitem[{Wilson \& Militzer(2014)}]{WilsonMilitzer2014}
---. 2014, Astrophys. J., 973, 34

\bibitem[{Wisdom(1996)}]{Wisdom1996}
Wisdom, J. 1996, {Non-perturbative Hydrostatic Equilibrium}.
\newblock \url{https://hdl.handle.net/1721.1/144248}

\bibitem[{Wisdom {et~al.}(2022)Wisdom, Dbouk, Militzer, Hubbard, Nimmo, Downey,
  \& French}]{Wisdom2022}
Wisdom, J., Dbouk, R., Militzer, B., {et~al.} 2022, Science, 377, 1285,
  \dodoi{10.1126/science.abn1234}

\bibitem[{Zeldovich \& Raizer(1968)}]{Ze66}
Zeldovich, Y.~B., \& Raizer, Y.~P. 1968, Elements of Gasdynamics and the
  Classical Theory of Shock Waves (New York: Academic Press)

\bibitem[{Zharkov \& Trubitsyn(1978)}]{ZT1978}
Zharkov, V.~N., \& Trubitsyn, V.~P. 1978, Physics of Planetary Interiors
  (Pachart), 388

\end{thebibliography}

\end{document}